\documentclass[preprint,aps, nofootinbib]{revtex4}
\input{epsf.tex}
\def\be{\begin{equation}}
\def\ee{\end{equation}}
\def\ba{\begin{eqnarray}}
\def\ea{\end{eqnarray}}

\begin{document}

\def\ba{\begin{eqnarray}}
\def\ea{\end{eqnarray}}
\def\be{\begin{equation}}
\def\ee{\end{equation}}
\def\tr{{\rm tr}}
\def\gtorder{\mathrel{\raise.3ex\hbox{$>$}\mkern-14mu
             \lower0.6ex\hbox{$\sim$}}}
\def\ltorder{\mathrel{\raise.3ex\hbox{$<$}\mkern-14mu
             \lower0.6ex\hbox{$\sim$}}}
\rightline{ORSAY LPT 04-16, DAMTP-2004-23}
\title{Models for the Brane-Bulk Interaction: Toward 
Understanding Braneworld Cosmological Perturbations}

\author{
Pierre Bin\'etruy,${}^{1,2}$ 
Martin Bucher,${}^{3,2,1}$ 
and Carla Carvalho${}^3$ \\
${}^1$ Laboratoire de Physique Th\'eorique, Universit\'e Paris-Sud, Orsay,
F-75231 France\\
${}^2$ APC, Universit\'e Paris VII (Denis-Diderot), 2 place Jussieu, 
75251 Cedex 05, France\\
${}^3$ DAMTP, Centre for Mathematical Sciences, University of Cambridge,\\
Wilberforce Road, Cambridge CB3 0WA, United Kingdom
}

\date{15 March 2004}

\begin{abstract}%
Using some simple toy models, we explore the nature of the brane-bulk
interaction for cosmological models with a large extra dimension. We
are in particular interested in understanding the role of the bulk
gravitons, which from the point of view of an observer on the brane
will appear to generate {\it dissipation} and {\it nonlocality},
effects which cannot be incorporated into an effective (3+1)-dimensional
Lagrangian field theoretic description. We explicitly work out the dynamics of
several discrete systems consisting of a finite number of degrees of freedom 
on the boundary coupled to a (1+1)-dimensional field theory
subject to a variety of wave equations. Systems both with and without
time translation invariance are considered and moving boundaries 
are discussed as well. The models considered contain all the qualitative
feature of quantized linearized cosmological perturbations for a 
Randall-Sundrum universe having an arbitrary expansion history, with the sole
exception of gravitational gauge invariance, which will be treated in
a later paper. 
\end{abstract}

\maketitle

\section{Introduction}

In this paper we consider the problem of computing the
cosmological perturbations for the 
one-brane Randall Sundrum model \cite{rs, langloisa}
in which we inhabit a (3+1)-dimensional brane
embedded in a pure (4+1)-dimensional anti de Sitter bulk spacetime. 
A $Z_2$ symmetry about the brane is supposed and the perturbations are treated
to linear order. The only bulk degrees of freedom are the (4+1)-dimensional
bulk gravitons or gravity waves, which propagate in the bulk \cite{rs, gt1}.
These are 
emitted, absorbed, and reflected by the brane as well as reflected, or 
perhaps, more properly, diffracted by the bulk, making the problem of 
predicting the cosmological perturbations in a braneworld scenario 
much more complex than its $(3+1)$-dimensional counterpart. 

A vast literature exists on the problem of calculating cosmological
perturbations in a braneworld scenario. 
Some progress has been made, either by resorting to various
approximations or by considering special spacetimes on the brane,
such as eternal pure de Sitter space, for which additional symmetry can
be exploited. A non-exhaustive sampling of the literature 
can be found in ref.~\cite{first_brane_pert}--\cite{last_brane_pert} 
and the references cited therein.
In this paper we develop some techniques that could be employed
to solve the problem completely, for a general expansion history
and without resort to any approximations. The work reported here, 
which is most closely related to considerations in 
Gorbunov, Rubakov and Sibiryakov \cite{rubakov} 
and the approach presented in the series of papers by Mukoyama 
\cite{mukoyama_first},
constitutes a first step toward this goal. 
Although we only discuss the connection with the Randall-Sundrum 
cosmologies, the methods and ideas developed here
would  apply equally well to more complex and perhaps more realistic
situations with additional degrees of freedom in the bulk.

It is this coupling between the brane and bulk degrees of freedom that
renders the problem of cosmological perturbations in the braneworld 
scenario difficult. For linearized cosmological perturbations in the 
ordinary (3+1)-dimensional cosmology, the problem may be {\it diagonalized}
by expanding the three spatial dimensions in Fourier components 
$\exp [i{\bf k}\cdot {\bf x}]$ (where for simplicity here we assume a 
spatially flat universe) \cite{cosmo_pert_rev}.
To linear order, there is no mixing between 
different ${\bf k}.$ Consequently, each such block may be analysed 
separately. The equations may be diagonalized further by
separating the {\it scalar,} {\it vector,} and {\it tensor} sectors.
Here we define any quantity expressible in terms of derivatives acting
on a scalar as {\it scalar}. Likewise, any quantity expressible
as derivatives acting on a pure {\it vector} potential is regarded
{\it vector,} etc. Under these definitions, spatial differentiation does not 
mix these three sectors, giving the evolution equations a block diagonal form.
For each {\bf k} there is  a finite number of degrees of freedom whose
time evolution is described by a finite number of ordinary coupled
differential equations in cosmic time $t.$ 

In the presence of an extra dimensional ``bulk,'' the number of degrees
of freedom in each {\bf k} sector is 
greatly---in fact, infinitely---enlarged. (See, for example, the review
\cite{durrer}). 
When the size of the extra dimension is finite, the additional bulk
degrees of freedom are discretely infinite, their spacing 
inversely proportional to the size of the extra dimension. 
When the extra dimension is very small, a large mass
gap must be overcome in order to access the infinite tower of bulk 
excitations and at low energies the (3+1)-dimensional limit is obtained. 
However, when the extra dimension is infinite, the bulk degrees
of freedom are continuous, labeled by an index $k_5,$ where
$0\le k_5<+\infty .$ For the case of a warped bulk spatial geometry, it 
has been shown how the coupling to the brane of the lowest energy
degrees of freedom is suppressed, giving on large scales 
ordinary (3+1)-dimensional gravity on the brane.  However, 
we would like to explore the corrections to such a limit. 
We would like to discover any possible differences in the 
predictions for the cosmological perturbations
between the standard (3+1)-dimensional cosmology and a
braneworld cosmology. Consequently, we must 
solve the interacting brane-bulk system.

The vast enlargement of the number of degrees of freedom 
renders the problem difficult in two respects: Firstly,
the equations become more complicated. But this is not all.
Secondly, a solution to the problem of braneworld cosmological
perturbations requires specifying initial conditions for the 
infinite number of extra degrees of freedom. While the first
problem is of a technical, computational character, the second
is of a more fundamental or physical nature. 

\begin{figure}
\begin{center}
\epsfxsize=6in \epsfysize=3in
\begin{picture}(300,200)
\put(-30,-10){$(a)$}
\put(95,-10){$(b)$}
\put(215,-10){$(c)$}
\put(330,-10){$(d)$}  
\put(-60,1){\leavevmode\epsfbox{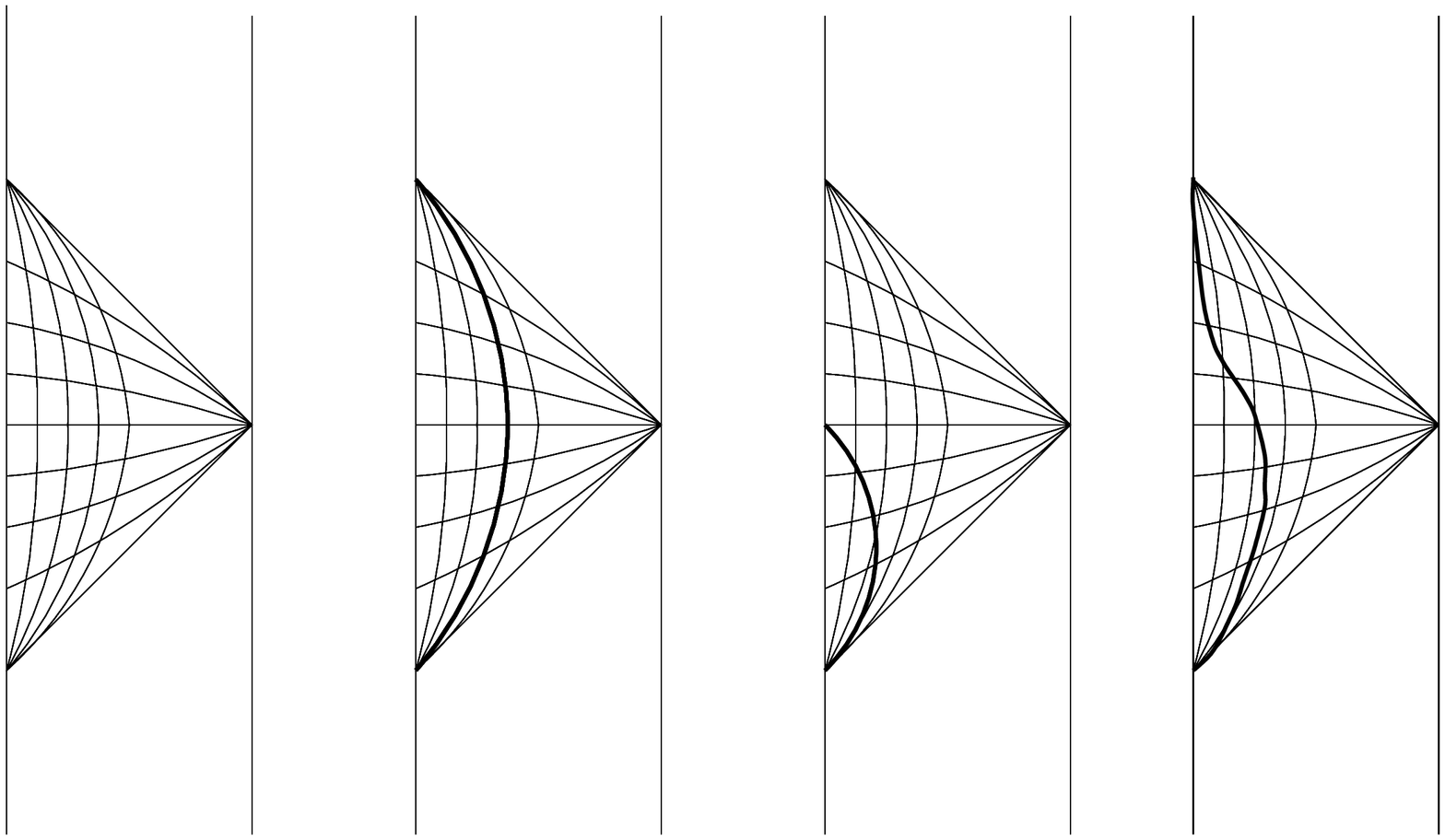}}
\end{picture} \end{center} \vskip 15pt
\caption{\small \baselineskip=10pt {
{\bf Various braneworld cosmologies embedded into the conformal
diagram for maximally-extended $AdS^5$.} Panel (a)
shows the triangular portion of maximally-extended AdS covered
by the Randall-Sundrum coordinates. Time runs upward and the
horizontal direction represents the fifth dimension, the other
three transverse dimensions being orthogonal to the page.
The family of lines focusing to the vertex on the right
represents surfaces of constant Randall-Sundrum time.
These asymptote in the future and in the
past to the null surfaces $H_{(+)}$ and $H_{(-)},$ which
are the future and past Cauchy horizons of this triangular
patch. The vertical curves connecting the lower
vertex to the upper vertex of the triangular regions represent
surfaces where the Randall-Sundrum coordinate for the fifth
dimension is constant. These are also surfaces on which the
scale factor $a(x_5)$ for the three transverse spatial dimensions
is constant. As one passes from right to left this scale factor
increases. In panel (b) the trajectory of the brane (indicated by
the heavy timelike curve) for a static Randall-Sundrum universe,
having a Minkowski induced geometry, is indicated. In panels
(c) and (d) we consider expanding universes, where the motion
of the brane to the left (with respect to the surfaces of constant
$x_5$) implies expansion. In panel (c) the brane worldline for a
de Sitter induced geometry on the brane is illustrated. The brane emanates
from the lower vertex, as for the static brane, but because
of the rapidity of the expansion, strikes the boundary at
conformal infinity before reaching the upper vertex. In
other words, de Sitter proper time on the brane becomes
infinite at finite Randall-Sundrum time. Panel (d)
illustrates a braneworld universe that is initially
inflating but whose expansion later slows down to
become a dust-dominated universe. In this case, because
of the deceleration, future infinity on the brane corresponds
to infinite Randall-Sundrum time. In all cases there
is a past Cauchy horizon $H_{(-)}$ on which initial data
for the bulk modes must be specified to completely
determine the subsequent evolution of the
coupled brane-bulk system.
}}
\label{Fig:ads}
\end{figure}

The problem of initial conditions for the bulk degrees of freedom
can be further elucidated by considering the Penrose diagram for
Randall-Sundrum Universes with various expansion histories, as indicated in 
Fig.~\ref{Fig:ads}. Panel (a) indicates the 
patch on the conformal diagram covered by the 
standard Randall-Sundrum coordinates,
having the line element
\ba
ds^2=dx_5^2+\exp [-2x_5/\ell ]\cdot \Bigl[ -dt^2+dx_1^2+dx_2^2+dx_3^2\Bigr] .
\ea
Panel (b) indicates a static $M^5$ brane
universe. Here the brane trajectory, indicated by the heavy curve,
coincides with one of the $x_5$ constant surfaces.
$H_{(+)}$ and $H_{(-)}$ are the bulk horizons with respect to the observers on
the brane. 
Panel (c) indicates a Randall-Sundrum universe with a $dS^5$ geometry
on the brane, and panel (d) indicates a universe that is initially
inflationary but then reheats to become a decelerating (e.g., dust-dominated)
universe.
In all cases, there is the 
past horizon $H_{(-)},$ of particular interest to us,
where initial data for the bulk gravitons must be specified. 

\begin{figure}
\begin{center}
\epsfxsize=6in \epsfysize=3in
\begin{picture}(300,200)
\put(-30,155){$H_{(+)}$}
\put(-30,55){$H_{(-)}$}
\put(0,-30){$(a)$}
\put(300,-30){$(b)$}
\put(-60,1){\leavevmode\epsfbox{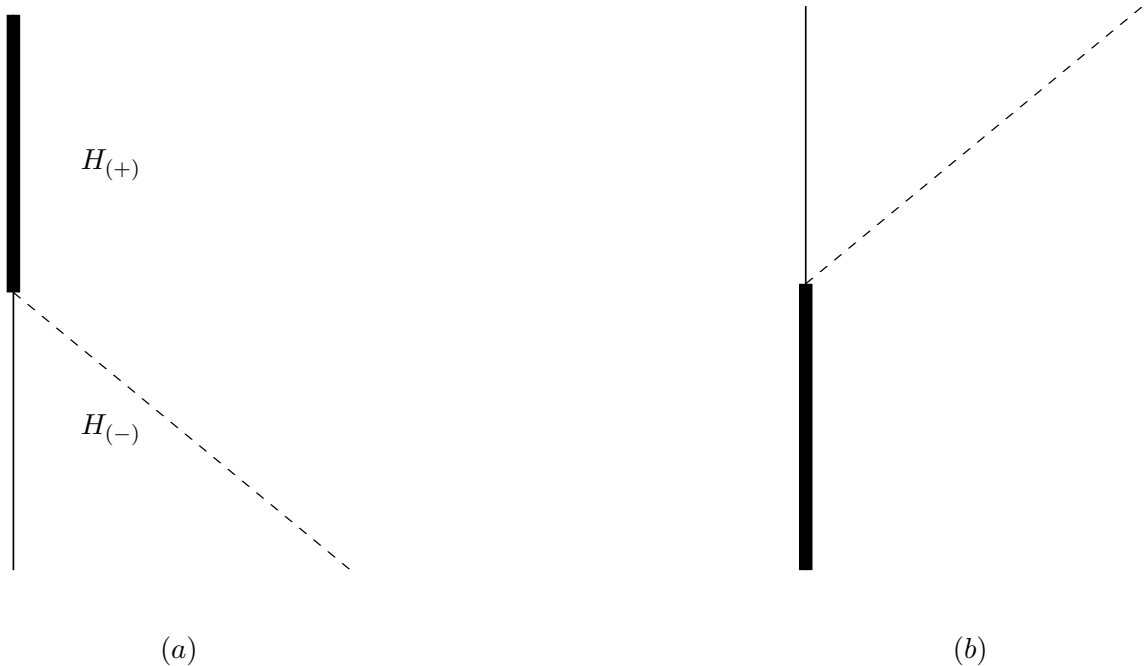}}
\end{picture} \end{center} \vskip 15pt
\caption{\small \baselineskip=10pt {
{\bf Fundamental processes for the bulk-brane interaction.}
The thin vertical solid line indicates the position of the 
brane and the bulk is situated to its right. A bulk graviton,
indicated by the dashed line, may strike the brane,
annihilating itself and creating an excitation of the brane,
indicated by the heavy vertical line, as illustrated in
panel (a). Likewise, an excitation of the brane may decay,
emitting a bulk graviton, as illustrated in
panel (b). Because we are interested here only
in the linearized theory, there are no vertices joining
more than two external lines. Propagation in the bulk may
involve scattering and is not in general along straight 
trajectories. Moreover, bulk gravitons may reflect off
the boundary without changing the excitations on the brane.
}}
\label{Fig:fund}
\end{figure}

An initial state for the brane-bulk system may be 
characterized completely by specifying the quantum
state of the incoming gravitons on $H_{(-)}$ and 
of the degrees of freedom on the brane at the intersection
of the brane with  $H_{(-)}.$ Subsequently, the bulk and
the brane degrees of freedom interact. The relevant 
fundamental processes are illustrated in Fig.~\ref{Fig:fund},
As shown in panel (a), bulk gravitons may be absorbed 
and transformed into 
quanta on the brane. Similarly, as shown in panel (b),
brane excitations may decay through the emission of 
bulk gravitons. These may either escape to future
infinity, leading to a sort of dissipation,
or be re-absorbed by the brane, leading to what 
from the four-dimensional point of view appears
as {\it nonlocality}.   Such {\it dissipation} and 
{\it nonlocality} are effects that cannot be 
incorporated into an effective four-dimensional
effective field theory description, and it is
hoped that by studying these phenomena one
might be able to find some distinctive observational
signature for the presence of extra dimensions.

In this article we present some simple toy models where a
single or finite number of degrees of freedom is coupled
to a one-dimensional continuum of degrees of freedom. Such
(1+1)-dimensional field theories coupled to a boundary 
having its own dynamics are the analogue of a fixed ${\bf k}$-sector
of the brane-bulk system. The degrees of freedom on the boundary
represent the degrees of freedom on the brane and the degrees
of freedom of the one-dimensional continuum represent the degrees
of freedom of the bulk. 
We consider a number of
examples of increasing complexity that exhibit all the 
qualitative features of the brane-bulk system except those concerning 
gravitational gauge invariance and gauge fixing. These
questions shall be treated in a future publication. 

In Section II we study the simplest such system. A harmonic oscillator
is coupled to a stretched string so that excitations of the oscillator
may decay, or dissipate, emitting waves on the string that propagate
to infinity. If there are no incoming waves on the string, which
is classically possible but not quantum mechanically, the 
motion of the oscillator is described exactly by the 
equation of motion for the dampened harmonic oscillator. 
Similarly, incoming waves may excite the oscillator.
In the analogy to braneworld cosmology, the oscillator
represents a degree of freedom on the brane and the
string represents the bulk gravitons. 
We work out the quantum mechanical description of this system
and in particular of its effective description in detail.
We show how the annihilation and creation operators localized
at the end
of the string, $a_{osc}(t)$ and $a_{osc}^\dagger (t),$
do not have definite frequency but rather are the superposition
of modes of different frequency having a certain characteristic
width of order the classical decay rate $\gamma .$ It is this spread
in frequency that causes the commutator 
$[a_{osc}(t),a_{osc}^\dagger (t')]$ to decay in modulus when 
$\vert t-t'\vert $ becomes large compared to $\gamma ^{-1}.$
We also consider several oscillators coupled to each other 
and to a string and the case where the mass density of
the string is not uniform so that there are reflections,
leading to nonlocality in the effective description for the 
oscillator degree of freedom. All the systems in this 
section have a time-translation invariance, so that
a Fourier decomposition in time can be employed,
greatly simplifying the problem. 

In Section III we consider cases where there is no
time translation invariance, in which the coupling
evolves with time. While the examples of the previous
section are analogous to the static Randall-Sundrum
brane, where the brane geometry is that of (3+1)-dimensional
Minkowski space, the examples of section III are more
akin to the expanding braneworld universe, where the 
couplings of the various modes to the brane evolve with
time. In the time translation invariant case, at any given
finite time, the quantum state of a mode depends only on
the quantum state in the bulk at past infinity. However,
with the time-dependent interactions, cases may be contemplated
where the state on the brane depends on a linear combination
of the initial state on the brane and that in the bulk. 
Some explicit examples are worked out where the linear
canonical transformation---a sort of $S$ matrix between 
the ``in'' and ``out'' states---is calculated. 
The matrix linking the ``in" and ``out" states
for the bulk is akin to the calculation in 
Gorbunov, Rubakov and Sibiryakov \cite{rubakov}
where the Bogoliubov transformation between the 
``in" and ``out" states is calculated for a brane
having a $dS^4$ geometry due to a pure cosmological
constant and nothing else on the brane. 

In section IV we consider the complications that arise when the 
wave equation on the string is such that the general solution can
no longer be decomposed into left and right movers. Any sort
of mass term or non-uniformity in the speed of propagation will 
render such a decomposition impossible by
causing disturbances emitted from the brane to scatter and propagate 
back onto the boundary. There can be many such multiple reflections.
In the case of interest to us, the fact that the bulk geometry 
is $AdS^5$ rather than $M^5$ causes such scatterings, which might
also be described as {\it diffraction,} particularly for wavelengths
of order the AdS curvature scale. We develop a perturbative 
approach to treating such scattering. The final result
indicates how the bulk interaction in the toy model
can be expressed in terms of integral kernels. 
The approach is similar to that sketched in the 
papers of Mukoyama \cite{mukoyama_first}.

Finally, in section V we consider the case where the 
boundary follows an arbitrary given timelike trajectory
$x(t).$ (Here $x_5$ is simply shortened to the 
$x$ of our toy models.)
We show how the relevant Green's functions may
be constructed using virtual sources whose strengths
are determined by solving a Volterra integral equation
of the second type. These techniques are relevant
to the brane-bulk problem because the propagation 
of the bulk gravitons is most trivial in a coordinate
system where the bulk is static but the brane moves
as a result of the expansion of the universe.  
In section VI we conclude with some 
closing comments.

\vfill\eject
\section{Examples With Time Translation Invariance}

\subsection{Quantum Mechanics With Dissipation : A Single Mass Coupled to
a String}

\begin{figure}
\begin{center}
\epsfxsize=3in
\epsfysize=3in
\begin{picture}(300,200)
\put(200,165){$u(t,x)$}
\put(150,165){$\mu$}
\put(275,175){$\tau$}
\put(85,165){$q(t)$}
\put(90,150){$m$}
\put(80,1){\leavevmode\epsfbox{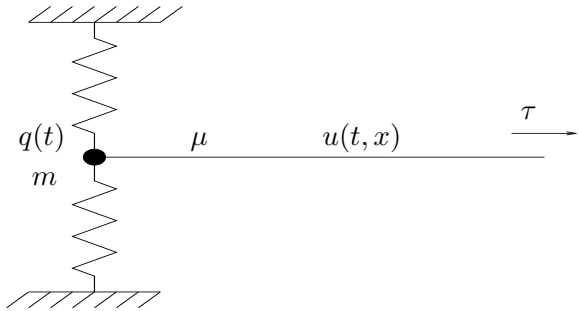}}
\end{picture}
\end{center}
\vskip -1.5in
\caption{\small \baselineskip=4pt {
{\bf Simple model for the brane-bulk interaction.}
 A simple harmonic oscillator consisting of a mass $m$
bound to a spring of stiffness $k=m\bar \omega ^2$ is
coupled to a stretched string of linear density $\mu $
and tension $\tau .$ $q(t)$ represents the vertical
displacement of the oscillator, and the field $u(t,x)$
represents the vertical displacement of the stretched
string. Here the oscillator represents
a degree of freedom localized on the brane and the
modes on the string represent the bulk gravitons.
As the mass oscillates, energy escapes into
the bulk, leading to dissipation. If the linear
density of the string is not uniform, some of
these modes scatter back, leading to effects
nonlocal in time.
}}
\label{Fig:spring}
\end{figure}

After the linearized perturbations of the brane-bulk system
have been decomposed into their Fourier components with
respect to three transverse spatial dimensions, the evolution
equations decompose into independent blocks, each labeled
by the three-dimensional wave number ${\bf k}.$ Each such
block has the dynamics of the $(1+1)$-dimensional field
theory coupled to one or more degrees of freedom localized
on the brane boundary. In this section, we study a simple
mechanical model of such a block. We consider a simple harmonic oscillator 
attached to the end of a string (see Figure \ref{Fig:spring}).
In the absence of coupling to 
the string, the equation of motion for the oscillator is
\be
m\ddot q(t)+m{\bar \omega }^2q(t)=F(t)
\label{osc}
\ee
where the external driving force $F(t)$ vanishes. The 
quantum states are simply the usual simple harmonic oscillator
ladder of states
of energy 
$E_n=\hbar \bar \omega (n+\frac{1}{2}).$

We now introduce dissipation in a `conservative' way---that is,
we couple the above oscillator to a system with an infinite
number of degrees of freedom in such a way that a classically
excited state of the oscillator progressively radiates its
energy into the continuum and this energy never returns.  
(Some other approaches to treating dissipation within the
framework of quantum mechanics may be found in refs. \cite{quant_dissip}.)
For concreteness, consider a string of linear density $\mu $ 
and tension $\tau .$ For simplicity, 
we set $\mu =\tau $ for the string, so that 
\be 
\frac{\partial ^2u}{\partial t^2}-
\frac{\partial ^2u}{\partial x^2}=0.
\ee 
If we attach the string to the oscillator at $x=0,$ the driving force 
exerted by the string on the mass becomes
\be 
F(t)=\tau \frac{\partial u}{\partial x}\Bigg| _{x=0}.  \ee 

Classically, we solve this system by postulating 
as an initial condition the absence
of any incoming (or left-moving) waves on the string.
Using the ansatz 
\be 
u(t,x)=C~\exp [i\omega (x-t)]
\ee
and setting $\gamma =\tau /m,$ we obtain from (\ref{osc}) where $q(t) = 
u(t,x=0)$ the quadratic equation 
\be
{\bar \omega }^2-\omega ^2=i\gamma \omega ,
\ee
having the solution
\be 
\omega =\frac{-i\gamma }{2}\pm \sqrt{\bar \omega ^2-{\gamma ^2\over 4}}. 
\label{omega}
\ee
This solution, of course, diverges at large positive $x$ 
and large negative $t;$ however, these divergences are 
unphysical, because in constructing a Green's function,
we match to a vanishing solution below a line of 
$(t-x)<t'$ where $t'$ is the time corresponding to 
the singular source located at $x=0$
at the oscillator.
Classically, from an effective point of view, we could
dispense with the string and its degrees of freedom infinite
in number and instead simply use the equation
\be 
m\Bigl[ \ddot q(t)+\gamma \dot q(t) +{\bar \omega }^2q(t)\Bigr] =
F_{ext}(t)
\label{eqn:a}
\ee
where the force exerted by the string on the oscillator at $x=0,$
$F_{string}(t),$ has been placed on the left-hand side and 
is now the dissipative term $m\gamma \dot q(t),$ and 
$F_{ext}(t)$ is included to denote an external force on 
the oscillator other than that exerted by the string.
The arrow of time here was set by our initial condition that
there be no incoming waves on the string in the past. If we 
had instead postulated no incoming waves in the future,
the sign of the $\gamma \dot q $ term would be inverted. 
It is important to note that, from a classical point of 
view, eqn.~(\ref{eqn:a}) is exact. There are no corrections. 

Quantum mechanically, the initial condition above does not
make sense. Vacuum fluctuations of the string 
require the presence of at least some left-moving waves.
Otherwise, it would not be possible to satisfy the 
canonical commutation relations. Consequently, we must proceed 
by some other means. We start by postulating a vacuum state
for the string-oscillator system. Note that when coupled to
the string in the above way, the oscillator degree of freedom
disappears. It has become part of the string, because
we have set $q(t)=u(t,x=0).$\footnote{%
To understand how this occurs, it is useful to 
consider the string as a continuum limit of a series
of small masses $(\Delta m)=\mu (\Delta l)$ connected by
springs of separation $(\Delta l)$ attached to
a harmonically bound mass of fixed magnitude $M$
at the end point. In the continuum limit, the 
frequency of the mode where the large mass 
oscillates relative to its neighbors approaches
infinity. It is always of energy of order the cutoff. 
}

For a fixed time dependence proportional to $e^{-i\omega t},$
we find the normal modes of the string with the ansatz
\be 
u_\omega (t,x)=\sin \Bigl[ \omega x-\phi (\omega )\Bigr] ~
e^{-i\omega t},
\ee
where $\omega$ is given in (\ref{omega}). We have
\be
\Bigl( {\bar \omega }^2-\omega ^2\Bigr) u_\omega(t,x=0)=\gamma 
\frac{\partial u_\omega}{\partial x}(t,x=0),
\ee
or equivalently
\be
\Bigl( {\bar \omega }^2-\omega ^2\Bigr) \sin \Bigl[\phi (\omega )\Bigr] =
-\gamma \omega \cos \Bigl[ \phi (\omega )\Bigr] ,
\ee
where $\gamma =(\tau /m).$ It follows that
\be 
\sin \Bigl[ \phi (\omega )\Bigr] =
\frac{ -\gamma \omega }{
\sqrt{
\gamma ^2\omega ^2+(\omega ^2-{\bar \omega }^2)^2
}
}
\approx    
\frac{
-(\gamma /2)}{
\sqrt{ (\gamma /2)^2+(\omega -\bar \omega )^2
}
},
\label{eqn:b}
\ee
where the latter approximation holds when $\gamma \ll \bar \omega ,$
in other words when the quality factor of the oscillator
is very large. 

The field on the string may be expanded as 
\be 
\hat u(t,x)=\int _0^{+\infty }\frac{d\omega }{\sqrt{2\pi \mu \omega }}
\sin \Bigl[ \omega x-\phi (\omega )\Bigr]
\Bigl[ \hat a(\omega )~e^{-i\omega t}+\hat a(\omega )^\dagger ~e^{+i\omega t}
\Bigr] 
\ee
where 
\be
[\hat a(\omega ), \hat a(\omega ')^\dagger ]=\delta (\omega -\omega ') .
\ee
The normalization factor follows from the canonical commutation relations
for the field $\hat u(t,x).$
It follows that the quantum operator corresponding to 
the oscillator degree of freedom localized
at the end of the string at time $t$ is given by 
\begin{eqnarray} 
\hat q(t)&=&-\int _0^{+\infty }\frac{d\omega }{\sqrt{2\pi \mu \omega }}
\sin \Bigl[\phi (\omega )\Bigr] 
\Bigl[ \hat a(\omega )~e^{-i\omega t}+\hat a(\omega )^\dagger ~e^{+i\omega t}
\Bigr] \nonumber\\
&=&\int _0^{+\infty }\frac{d\omega }{\sqrt{2\pi m \gamma \omega }}
\frac{
\gamma \omega
}{
\sqrt{
\gamma ^2\omega ^2+(\omega ^2-{\bar \omega }^2)^2
}
}
\Bigl[ \hat a(\omega )~e^{-i\omega t}+\hat a(\omega )^\dagger ~e^{+i\omega t}
\Bigr] .
\end{eqnarray}
Define
\be
\rho ^{1/2}(\omega )
=N~\sqrt{\frac{\gamma \omega }{\gamma ^2\omega ^2+(\omega ^2
-{\bar \omega }^2)^2}}
\ee
where
\be 
N^{-2}=\int _0^{+\infty }d\omega ~\frac{\gamma \omega }
{\gamma ^2\omega ^2+(\omega ^2 -{\bar \omega }^2)^2}.
\ee
We may define 
\be
\hat a_{osc}(t)=\int _0^{+\infty }d\omega ~\rho ^{1/2}(\omega )~
\exp [-i\omega t]~\hat a(\omega ).
\ee
Because
\be 
\int _0^{+\infty }d\omega ~\rho (\omega )=1,
\ee
it follows that at equal times
\be 
\left[ \hat a_{osc}(t), {\hat a_{osc}}^\dagger (t)\right] =1,
\ee
as for an ordinary harmonic oscillator. 
However, in the general case
\be 
\left[ \hat a_{osc}(t), {\hat a_{osc}}^\dagger (t')\right] 
=\int _0^{+\infty }d\omega ~\rho (\omega )~\exp [-i\omega (t-t')],
\ee
which for unequal times is of a modulus smaller than one.
For $\vert t-t'\vert \ll \gamma ^{-1}$ this is to a good 
approximation a simple phase, namely $\exp [-i\bar \omega (t-t')],$
just as one would obtain by ignoring the dampening and pretending
that the oscillator is uncoupled with respect to the string.
However, when $\vert t-t'\vert $ becomes comparable to
or greater than $\gamma ^{-1},$ one observes a sizable 
diminution in modulus. Unless the density $\rho (\omega )$
has some delta functions spikes, 
the Riemann-Lebesgue lemma implies that, for times 
sufficiently
well separated, the operators almost commute.

Let us consider the special case of a uniform string of
small tension, so that the approximation on the far right-hand
side of eqn.~(\ref{eqn:b}) holds where $\gamma \ll \bar \omega .$ In this
case 
\be 
\rho (\omega )=
\frac{1}{\pi }
\frac{(\gamma /2)}{(\gamma /2)^2 +(\omega -\bar \omega )^2}
\ee 
and with very little error we may replace the interval of
integration $[0,+\infty )$ with the doubly-infinite interval
$(-\infty , +\infty ),$ so that
\begin{eqnarray}
\left[ \hat a_{osc}(t), {\hat a_{osc}}^\dagger (t)\right]
&\approx &\frac{1}{\pi }
\int _{-\infty }^{+\infty }d\omega ~
\frac{(\gamma /2)}{(\gamma /2)^2 +(\omega -\bar \omega )^2}~
\exp [-i\omega (t-t')]\nonumber\\
&=& \exp [-i\bar \omega (t-t')]\cdot \exp [-(\gamma /2)\vert t-t'\vert ].
\end{eqnarray}
For broad resonances, there are corrections to this formula,
but the general qualitative behavior persists.

\subsection{Multiple reflections and non-locality}

We may also contemplate more complicated systems for which the 
classical behavior of the oscillator is no longer local in time,
as for example as indicated in eqn.~(\ref{eqn:a}), but rather an equation
of the form 
\be
\ddot q(t)+ \gamma \dot q(t)+
{\bar \omega }^2 q(t)=\frac{1}{m}\left[
F_{ext}(t)+
\int _{-\infty }^t dt'~G(t-t')~F_{ext}(t')
\right] 
\ee
where the kernel $G(t-t')$ represents the amplitude for a 
wave on the string generated at time $t'$ reflecting back
to strike the oscillator subsequently at time $t.$ 
Such nonlocal behavior will arise in the brane-bulk system
because bulk gravitons initially emitted by the brane 
will be scattered back to the brane by the curved bulk geometry. 

For concreteness, consider the case where rather than being 
uniform, the linear density of the string abruptly increases
at $x=L,$ so that at that point the propagation speed suddenly drops
from $c_l$ to $c_r$ as one passes from the left to the right.
Physically, a portion of the outgoing
wave is transmitted and another is reflected back toward the 
oscillator. Across the junction from left to right, we have 
the following matching rules (which follows from
continuity of the amplitude and its first 
derivative)
\begin{eqnarray}
\cos \left[ \omega \frac{(x-L)}{c_l} \right] 
&\to &
\cos \left[ \omega \frac{(x-L)}{c_r} \right] ,
\nonumber\\
\sin \left[ \omega \frac{(x-L)}{c_l} \right]
&\to &
\left( \frac{c_r}{c_l} \right) 
\sin \left[ \omega \frac{(x-L)}{c_r} \right] ,
\end{eqnarray}
so that the solution for $x\le L$ satisfying the 
boundary condition at the oscillator 
\be
\sin [\omega x/c_l+\phi (\omega )]
=\sin [\omega (x-L)/c_l]~\cos [\phi (\omega )+\omega L/c_l]+
 \cos [\omega (x-L)/c_l]~\sin [\phi (\omega )+\omega L/c_l],
\ee
for $x\ge L$ transforms into 
\be
(c_r/c_l)\cdot 
\sin [\omega (x-L)/c_r]~\cos [\phi (\omega )+\omega L/c_l]+
\cos [\omega (x-L)/c_r]~\sin [\phi (\omega )+\omega L/c_l],
\ee
which has the overall amplitude 
\be
A=\sqrt{ (c_r/c_l)^2~\cos ^2[\phi (\omega )+\omega L/c_l]
                    +\sin ^2[\phi (\omega )+\omega L/c_l]
}.
\ee
Since the string is semi-infinite, it is this amplitude alone that
determines the normalization of the modes on the string. 
The amplitude on the interval $[L,+\infty )$ infinitely
outweighs that on $[0,L].$
It follows that the spectral density takes the form 
\be 
\rho (\omega )=\bar N \frac{\sin ^2[\phi (\omega )]}%
{ (c_r/c_l)^2~\cos ^2[\phi (\omega )+\omega L/c_l]
                    +\sin ^2[\phi (\omega )+\omega L/c_l]
}
\label{mk:a}
\ee
where $\bar N$ is a normalization constant. 
Qualitatively, the density has the form of an ordinary
resonance (from the numerator) masked by a function 
with spikes at $\phi (\omega )+\omega L/c_l=n\pi $
of a certain width. The closer the reflection coefficient
$R=(c_l-c_r)/(c_l+c_r)$ is to unity, the sharper the spikes.
In the extreme limit $c_r\to 0,$ the reflection is total, 
so that the system is effectively no longer 
semi-infinite but rather a finite cavity, with a discrete
spectrum. In this case the spikes have a $\delta $-function
character. The presence of these periodically spaced 
spikes may be understood as follows. The
normalization of the modes of a given $\omega$ is determined 
by the amplitude of the wave at $x>L.$ However, it is
only near a resonance that this wave has a significant
penetration into the region $x<L$ to the left of the 
junction. 

Consider the Fourier transform of 
\ba
f(T)=
\sum _{n=-\infty }^{+\infty }
R^{\vert n\vert }
\exp [i\delta n]
\exp [-i\bar \omega (T-2nL)]
\exp \left[ -{\gamma \over 2}\vert T-2nL/c_l\vert \right] , 
\label{rf:bb}
\ea
which represents the effect on the commutator 
of (multiple) reflections, with the real reflection
coefficient $R$ and phase shift $\delta $ describing
the effect of a single reflection. 
This Fourier transform is simply
\ba
\rho   (\omega )=
\rho _0(\omega )
\sum _{n=-\infty }^{+\infty }
R^{\vert n\vert }
\exp [in\delta ]
\exp [ +2in\omega L/c_l]
\ea
where $\rho _0(\omega )$ is the Fourier transform of 
$\exp [-i\bar \omega T]\exp \Bigl[ -(\gamma /2)\vert T\vert \Bigr] .$
We may evaluate 
\ba
\sum _{n=-\infty }^{+\infty }
R^{\vert n\vert }
\exp [in\delta ]
\exp [ +2in\omega L/c_l]
=
{1-R^2\over 1+R^2-R
\Bigl( \exp [+i(2\omega L/c_l +\delta )]
+\exp [-i(2\omega L/c_l +\delta )]\Bigr) }.
\ea
Similarly, using the fact that $\phi (\omega )\approx \pi/2$
in the neighborhood of the resonance, we may rewrite the 
denominator in eqn.~(\ref{mk:a})
as 
\ba
&&
{ 1\over (c_r/c_l)^2~\cos ^2[\phi (\omega )+\omega L/c_l]
                    +\sin ^2[\phi (\omega )+\omega L/c_l] }
\approx 
{ 1\over (c_r/c_l)^2~\sin ^2[\omega L/c_l]
                    +\cos ^2[\omega L/c_l] }\cr 
&=&{2\over 
\left( 1+{c_r^2/c_l^2}\right) +
\left( 1-{c_r^2/c_l^2}\right) \cos [2\omega L/c_l] }\cr
&\propto &
{1\over 1+\lambda \cos [2\omega L/c_l]}
\label{rf:aa}
\ea
where $\lambda = (c_l^2-c_r^2)/(c_l^2+c_r^2)=2R/(1+R^2).$
We observe that 
eqn.~(\ref{rf:aa})
has the form of 
eqn.~(\ref{rf:bb}) 
with phase shift $\delta =\pi.$ 

The above example illustrates how for models having 
a time translation symmetry nonlocal effects are encoded
in the spectral density $\rho (\omega ).$ Here the 
multiple reflections alter the spectral density
$\rho _0(\omega )$
in the absence of reflections 
through multiplication by the mask function in
eqn.~(\ref{rf:aa}) having the profile of a picket 
fence. After an integer number of reflection times
$2L$ the phases of the various pickets interfere 
constructively causing a peak in the commutator.

We may also contemplate examples where 
the linear density of the string
$\mu (x)$ varies smoothly so that there is an
amplitude for reflection everywhere and these reflections
interfere with each other to give a rather complicated
response kernel $G(t-t').$ The method of calculating this
nonlocal kernel for the brane-bulk system shall be the 
subject of sections IV and V. 

\subsection{Interpretation in terms of Hilbert space}

Let us consider the physical interpretation of this diminution in
magnitude of the commutator. Our Hilbert space ${\cal H}_{string}$ 
has the structure of 
a direct product of an infinite number of harmonic oscillator
Hilbert spaces. There exists an infinite number of particle species.
In an abstract sense, let $f(\omega )$ be a 
normalized square-integrable complex function, that is
\be
\int _0^{+\infty }d\omega ~\vert f(\omega )\vert ^2 =1.
\ee
Each such function specifies a type of particle
whose annihilation operator is given by
\be
\hat a_f=\int _0^{+\infty }d\omega ~f(\omega )~\hat a(\omega ),
\ee
which together with its conjugate $\hat a_f^\dagger $ generates
a single-particle subspace 
${\cal H}_f \subset {\cal H}_{string}$.
We may also consider the Hilbert space of ``particle species'' (distinct
from the above spaces) endowed with a 
natural inner product: 
\be
\langle f\vert g\rangle =\int _0^{+\infty }d\omega ~f^*(\omega )~g(\omega ).
\ee
More precisely, particle species are labeled by complex rays in this Hilbert
space: any two such complex-valued, square-integrable
functions differing only by a complex phase represent the 
same kind of particle. 

Without coupling to the string, the type of particle coupled to 
the oscillator does not evolve with time. $\hat a(t)=\exp [-i\bar \omega t]
\hat a(t=0),$ and as noted a difference in phase does not correspond 
to a difference in particle type. 
However, in the case of coupling to the string,
the oscillator annihilation operator, because it is a superposition
of string modes of differing frequency, evolves in this ray space
of different particle species. Any quantum operator 
${\cal O}(t)$ localized at the oscillator at the end of the string
at a time $t$ may be expressed as a sum of products of $\hat a_{osc}(t)$ and 
$\hat a_{osc}^\dagger (t).$  
Any operator that cannot be so constructed is not localized at
the end of the string.

Physically, the fact that for large $t,$ $\hat a_{osc}(t)$ almost commutes with
$\hat a_{osc}(t=0)$ reflects the fact that the particles created by 
$\hat a_{osc}^\dagger (t)$ escape from the neighborhood of the end of 
the string and propagate down the string, eventually leaving no trace 
detectable by an observer able only to operate on the string end. 
Once the particle has escaped from the end of the string, no operator
localized at the string end is capable of altering its state.

Quantum mechanically, from the point of view of a quantum observer
only coupled to the mass at the end of the string,
calculating probabilities that involve ``dissipation''---that is, 
the escape of energy, particles, and information down the string---requires
summing over all such final states. That is, rather than considering
vacuum-to-vacuum amplitudes
we must sum over the particles not seen that escape down the string.

Let us first consider the simplest possible process, the creation of 
a single oscillator quantum at $t=t_i$ and the outcome of an attempt
to recover this particle at a subsequent time $t_f.$ One possible
outcome is that the particle is detected at $t=t_f.$ The amplitude
for this process is 
\be
{\cal A}=
\langle 0\vert 
\hat a_{osc}(t_f)~
\hat a_{osc}^\dagger (t_i)
\vert 0\rangle 
=
\exp[-i\bar \omega (t_f-t_i)] \exp[-(\gamma /2)(t_f-t_i)].
\ee
Note that the probability 
\be
p_{1_f\leftarrow 1_i}=\vert {\cal A}\vert ^2
=\exp[-\gamma (t_f-t_i)]
\ee
is strictly less than one, and 
\be
p_{0_f\leftarrow 1_i} =1-p_{1_f\leftarrow 1_i}=
1-\exp[-\gamma (t_f-t_i)] 
\ee
gives the probability that no particle is detected at the end 
of the string. This latter probability is actually the sum
of the probabilities for an infinite number of orthogonal
processes. One takes the infinite sum over $I$ where the index $I$
labels all the particle types orthogonal to the type coupled to
the string at $t=t_f.$ In other words, 
\be 
p_{0_f\leftarrow 1_i} = \sum _I \vert {\cal A}_{I_f\leftarrow 1_i}\vert ^2.
\ee
More complicated `dissipative' processes may be calculated 
analogously. 

\section{Examples Without Time Translation Invariance}

In the previous section we considered harmonic oscillators coupled
to a string. Our principal tool was the decomposition into Fourier
modes of fixed frequency with respect to time. To quantize we 
associated the positive frequency modes with the annihilation 
operators of the unique preferred vacuum state. In this 
section we generalize to situations where a global time translation
symmetry is lacking. We are in particular interested in cases 
where the coefficients characterizing the coupling of the 
oscillator or oscillators to the continuum evolve with time.
In this case Fourier methods cannot be used. Instead it is 
necessary to work in real space. In a braneworld cosmology
the expansion of the universe on the brane renders the system
time dependent. 

For the most general case, it is not possible to single out a 
unique preferred vacuum. However for a massless wave equation, 
for which any solution may be decomposed into left-moving
and right-moving waves, this is not a problem because 
it suffices to postulate
the usual vacuum state for the incoming waves. In other words,
where $\omega >0$ the modes having the form 
$\exp [-i\omega (t+x)]$ in the limit $t\to -\infty $ 
are associated with annihilation operators and the  
$\exp [+i\omega (t+x)]$ modes with the creation operators. 
This defines a unique vacuum state, even when the couplings 
to the harmonic oscillators at the end of the string 
and of the harmonic oscillators themselves vary with time.
The same applies when the equation becomes massless sufficiently
quickly as $x\to \infty .$ With respect to the oscillators, 
it is necessary to define an initial quantum state if the oscillators
decouple from the string sufficiently quickly as $t\to -\infty .$
If they do not decouple rapidly enough, 
the initial state becomes erased because the initial
state of the string degrees of freedom are infinitely more
relevant. 

In the first worked example we consider a single oscillator
whose spring constant, which is proportional to $\omega ^2(t),$
varies with time. If $\omega ^2(t)$ approaches a constant
as $\vert t\vert $ becomes large, there are well-defined 
``in" and ``out'' vacua. However, these vacua do not coincide
with each other.
The time variation in $\omega ^2(t)$ excites quanta through
parametric resonance. The production of quanta may be 
characterized by a Bogoliubov (i.e., linear canonical) 
transformation relating the modes of positive and negative 
frequency for the ``in'' and the ``out'' vacua. In this worked
example we consider the special case where  $\omega ^2(t)$
has the form of a step function. In this case, the applicable
integrals may be evaluated explicitly. However, the generalization
to an arbitrary form for $\omega ^2(t)$ approaching a constant for
large $\vert t\vert $ is straightforward. 

The second worked example is analogous except that rather than one
there are two harmonic oscillator degrees of freedom at the end
of the string. In this example, for $\vert t\vert >T$ the two
harmonic oscillators are uncoupled. Consequently, to define an initial
state, in addition to requiring the absence of incoming waves
it is necessary to characterize the ground state of the oscillator
initially uncoupled to the string. The resulting Bogoliubov
transformation from ``in'' to ``out'' states acts on a Hilbert space
with one discrete degree of freedom and a continuum of string degrees
of freedom. The generalization to many oscillators or to more
complicated time dependences is straightforward. 

\subsection{A single harmonic oscillator with time varying spring constant}

We return to example of section IIA where the stiffness of the oscillator
$\bar \omega ^2,$ formerly constant, now becomes a time-dependent function
$\bar \omega ^2(t).$ 
We consider the simplest system for which nontrivial Bogoliubov coefficients
may be calculated explicitly. 
Concretely, consider a harmonic oscillator of mass $m$ whose coupling
to a string of tension $\tau$ is turned on for only a finite interval
of time. The equation of motion is 
\ba
m\left[ 
{~d^2\over dt^2}+{\bar \omega}^2(t) 
\right] 
q(t)=F(t)
\ea
where the driving force $F(t)$ exerted by the string on the oscillator
is given by 
\ba
F(t)=\tau \frac{\partial u}{\partial x}\Bigg|_{x=0}. 
\ea
We may decompose the string solution, which obeys the massless
wave equation, into the two components
\ba
u(t,x)=u_{out}(t-x)+u_{in}(t+x).
\ea
The solution for the oscillator is simply
\ba
q(t)=u_{out}(t)+u_{in}(t).
\ea
It follows that
\ba
F(t)=\tau \frac{\partial u}{\partial x}\Bigg|_{x=0} 
=\tau \left( 
\frac{\partial u_{in}(t)}{\partial t} 
- \frac{\partial u_{out}(t)}{\partial t}
\right).
\ea
With $\gamma =(\tau /m)$, the equation of motion takes the form
\ba
\left[ 
{~d^2\over dt^2}+\gamma {~d\over dt}+\bar \omega ^2(t)
\right] 
u_{out}(t)
=
(-)\left[ 
{~d^2\over dt^2}-\gamma {~d\over dt}+\bar \omega ^2(t)
\right]
u_{in}(t)
\ea
or more symbolically
\ba
D_{out}(t)~u_{out}(t)=(-)D_{in}(t)~u_{in}(t).
\ea
Formally,
\ba
u_{out}(t)=(-)\int _{-\infty }^tdt'~G_{out}(t ~\vert ~t')~D_{in}(t')~u_{in}(t')
\ea
where $G_{out}$ is the retarded form of $D_{out}^{-1}.$ 
This means that $G_{out}$ mediates the propagation from $t'$ to $t$ of
the state $u_{in}$ subject to the potential $D_{in}$.

Consider the inverse Fourier transform of $u_{out}$, which can be
interpreted as the description of the state $u_{out}$ in the
Heisenberg representation. We obtain 
\ba
u_{out}(\omega_f) 
&=& {-1\over 2\pi}
\int_{-\infty}^{+\infty}dt_f~e^{+i\omega_f t_f}~
\int_{-\infty }^{t_f}dt_i~G_{out}(t_f ~\vert ~t_i)~D_{in}(t_i)~
\int_{-\infty}^{+\infty}d\omega_i~e^{-i\omega_i t_i}~ u_{in}(\omega_i)
\cr
&=& \int_{-\infty}^{+\infty}d\omega_i~S(\omega _f\vert \omega _i)~
u_{in}(\omega_i)
\ea
where 
\ba
S(\omega _f\vert \omega _i)={-1\over 2\pi}
\int _{-\infty }^{+\infty }dt_f~e^{+i\omega _ft_f}~
\int _{-\infty }^{t_f}dt_i~
G_{out}(t_f ~\vert ~t_i)~D_{in}(t_i)~
e^{-i\omega _it_i}
\ea
is the corresponding propagator in momentum space describing the
amplitude that a quantum with frequency $\omega_i$ be
later observed to have frequency $\omega_f.$ 
If we associate $e^{-i\omega t}$ with the annihilation operator, then 
$S(\omega _f\vert \omega _i)$ describes the probability amplitude 
that a quantum of frequency $\omega_i$ be destroyed and a quantum of
frequency $\omega_f$ be created. There are no homogeneous solutions
for $u_{out}.$ $u_{out}$ is completely determined by $u_{in}.$

In order to work out a concrete example, we set 
\ba
\bar \omega ^2(t)=
\left\{
\begin{array}{r@{\quad {\rm if}\quad}l}
(\gamma /2) ^2 +\Omega ^2, & \vert t\vert >T,\\
(\gamma /2) ^2 +\tilde \Omega ^2, & \vert t\vert <T.
\end{array}
\right.
\ea
It follows that the homogeneous equation $D_{out}(t)~u(t)=0$ has the solutions
\ba
u(t)=
\left\{
\begin{array}{r@{\quad {\rm for}\quad}l}
A~\exp [(-\gamma/2 +i\Omega )t]+B~\exp [(-\gamma/2 -i\Omega )t], 
& \vert t\vert >T,\\
C~\exp [(-\gamma/2 +i\tilde \Omega )t]+D~\exp [(-\gamma/2 -i\tilde
  \Omega )t], 
& \vert t\vert <T.
\end{array}
\right.
\ea

We adopt a sort of Heisenberg representation where, here for $\vert t\vert >T,$
the coefficients of the two solutions are expressed with respect to an
arbitrary origin at $t=t'$ (or $t=t^{\prime \prime }$):
\ba
u(t)&=&A_{t^{\prime }}~e^{(-\gamma/2 +i\Omega )(t-t^{\prime })}
      +B_{t^{\prime }}~e^{(-\gamma/2 -i\Omega )(t-t^{\prime })}\nonumber\\
     &=&A_{t^{\prime \prime }}~e^{(-\gamma/2 +i\Omega )(t-t^{\prime \prime })}
      +B_{t^{\prime \prime }}~e^{(-\gamma/2 -i\Omega )(t-t^{\prime \prime })}.
\label{our_not_a}
\ea
It follows that a change of origin is effected by the transformation
\be
\pmatrix{ A\cr B\cr }_{t^{\prime \prime }}\equiv 
\pmatrix{ A_{t^{\prime \prime }}\cr B_{t^{\prime \prime }}\cr }
={\bf P}(t^{\prime \prime }~\vert ~t^{\prime })
\pmatrix{ A\cr B\cr }_{t^{\prime }}
=
\pmatrix{ 
e^{(-\gamma/2 +i\Omega )(t^{\prime \prime }-t^{\prime })}&0\cr
0& e^{(-\gamma/2 -i\Omega )(t^{\prime \prime }-t^{\prime })}\cr
}
\pmatrix{ A\cr B\cr }_{t^{\prime }}.
\label{our_not_b}
\ee
Similarly, for $\vert t\vert <T,$
\ba
u(t)&=&C_{t^{\prime }}~e^{(-\gamma/2 +i\tilde \Omega )(t-t^{\prime })}
      +D_{t^{\prime }}~e^{(-\gamma/2 -i\tilde \Omega )(t-t^{\prime })}
\ea
and a change of origin is effected by the transformation
\be
\pmatrix{ C\cr D\cr }_{t^{\prime \prime }}=
\tilde {\bf P}(t^{\prime \prime }~\vert ~t^{\prime })
\pmatrix{ C\cr D\cr }_{t^{\prime }}
=
\pmatrix{
e^{(-\gamma/2 +i\tilde \Omega )(t^{\prime \prime }-t^{\prime })}&0\cr
0& e^{(-\gamma/2 -i\tilde \Omega )(t^{\prime \prime }-t^{\prime })}\cr
}
\pmatrix{ C\cr D\cr }_{t^{\prime }}.
\ee
Matching at a jump in $\bar \omega ^2$, 
as $(\tilde \Omega \leftarrow \Omega)$, 
is effected by the transformation:
\be
\pmatrix{ C\cr D\cr }_{t+\epsilon}
={\bf J}\pmatrix{ A\cr B\cr }_{t-\epsilon}
={1\over 2\tilde \Omega }\pmatrix{
(\tilde \Omega +\Omega )&
(\tilde \Omega -\Omega )\cr 
(\tilde \Omega -\Omega )&
(\tilde \Omega +\Omega )\cr 
}
\pmatrix{ A\cr B\cr }_{t-\epsilon},
\ee
and similarly, in the other direction, as $(\Omega \leftarrow \tilde \Omega)$,
\be
\pmatrix{ A\cr B\cr }_{t+\epsilon}
={\bf J}^{-1}\pmatrix{ C\cr D\cr }_{t-\epsilon}
={1\over 2\Omega }\pmatrix{
(\Omega +\tilde \Omega )&
(\Omega -\tilde \Omega )\cr 
(\Omega -\tilde \Omega )&
(\Omega +\tilde \Omega )\cr 
}
\pmatrix{ C\cr D\cr }_{t-\epsilon}.
\ee

In order to construct Green's functions, it is useful 
to divide the 
the real line into three regions
\ba
{\rm region ~I  }&=& (-\infty , -T],\cr
{\rm region ~II} &=& (-T , +T),\cr
{\rm region ~III}&=& [+T, +\infty ).
\ea

A Green's function for the oscillator 
where both endpoints lie within region
I (i.e., $t_i<t_f<-T$), is expressed in terms of the 
conventional notation as
\ba 
G(t,t_i)=(\partial_t^2+\gamma \partial_t+\omega ^2)^{-1}(t,t_i)=
{\sin [\Omega (t-t_i)]\over \Omega }\exp [-(\gamma/2) (t-t_i)]~
\theta(t-t_i).
\ea
In terms of our notation, where the homogeneous 
solutions are expressed according to the 
conventions of eqns.~(\ref{our_not_a}) and (\ref{our_not_b}),
the Green's function is generated by the column vector
\ba
\pmatrix{ {+1\over 2i\Omega }\cr {-1\over 2i\Omega }}
\label{ginsert}
\ea
localized at $t=t_i$ and placed at the far right in matrix expressions
time-ordered from right to left. It follows that,
in terms of our matrix notation, the Green's function is re-expressed as
\ba
G(t_f~\vert t_i)=
\pmatrix{1\cr 1\cr }^T
{\bf P}(t_f~\vert ~t_i)~
\pmatrix{ {+1\over 2i\Omega }\cr {-1\over 2i\Omega }}.
\ea
For propagation from region I into region II one instead would 
have
\ba
G(t_f~\vert t_i)=
\pmatrix{1\cr 1\cr }^T
\tilde {\bf P}(t_f~\vert -T)~
{\bf J}~
{\bf P}(-T~\vert  t_i)~
\pmatrix{ {+1\over 2i\Omega }\cr {-1\over 2i\Omega }}.
\ea
and the other cases may be worked out analogously in a 
straightforward manner. If $t_i$ lies in regions I or III,
the column vector (\ref{ginsert}) is used; otherwise,
if $t_i$ lies in region II, $\Omega $ in eqn. (\ref{ginsert})
is replaced with $\tilde \Omega .$

We now proceed to express the S matrix as the following sum 
\ba 
S(\omega _f \vert \omega _i)&=&
S_{I\leftarrow I}(\omega _f \vert \omega _i)+
S_{II\leftarrow I}(\omega _f \vert \omega _i)+
S_{III\leftarrow I}(\omega _f \vert \omega _i)\cr
&+&S_{II\leftarrow II}(\omega _f \vert \omega _i)+
S_{III\leftarrow II}(\omega _f \vert \omega _i)+
S_{III\leftarrow III}(\omega _f \vert \omega _i),
\ea
whose six terms are given by:
\ba
S_{I\leftarrow I}(\omega _f \vert \omega _i)&=&
{1\over 2\pi }
(-)\left[ {\gamma ^2\over 4}+\Omega ^2-\omega _i^2+i\gamma \omega _i~ 
\right]
\int _{-\infty }^{-T}dt_f~\int _{-\infty}^{t_f}dt_i~
e^{+i\omega _ft_f}
e^{-i\omega _it_i}
\cr  
&&\times 
\pmatrix{1\cr 1\cr }^T
{\bf P}(t_f \vert ~t_i)
\pmatrix{
{+1\over 2i\Omega }\cr 
{-1\over 2i\Omega }\cr 
}.
\ea
\ba
S_{II\leftarrow I}(\omega _f \vert \omega _i)&=&
{1\over 2\pi }
(-)\left[ {\gamma ^2 \over 4}+\Omega ^2-\omega _i^2+i\gamma \omega _i~
\right]
\int _{-T}^{+T}dt_f~\int _{-\infty }^{-T}dt_i~
e^{+i\omega _ft_f}
e^{-i\omega _it_i}
\cr  
&&\times 
\pmatrix{1\cr 1\cr }^T
\tilde {\bf P}(t_f \vert -T)~
{\bf J}~
{\bf P}(-T \vert ~t_i)
\pmatrix{
{+1\over 2i\Omega }\cr
{-1\over 2i\Omega }\cr
}.
\ea
\ba
S_{III\leftarrow I}(\omega _f \vert \omega _i)&=&
{1\over 2\pi }
(-)\left[{\gamma ^2\over 4}+\Omega ^2-\omega _i^2+i\gamma \omega _i~
\right]
\int _{+T}^{+\infty }dt_f~\int _{-\infty }^{-T}dt_i~
e^{+i\omega _ft_f}
e^{-i\omega _it_i}\cr 
&&\times 
\pmatrix{1\cr 1\cr }^T
{\bf P}(t_f \vert +T)~
{\bf J}^{-1}~
\tilde {\bf P}(+T~\vert -T)~
{\bf J}~
{\bf P}(-T~\vert ~t_i)
\pmatrix{
{+1\over 2i\Omega }\cr
{-1\over 2i\Omega }\cr
}.
\ea
\ba
S_{II\leftarrow II}(\omega _f \vert \omega _i)&=&
{1\over 2\pi }
(-)\left[ {\gamma ^2\over 4}+\tilde \Omega ^2-\omega _i^2+i\gamma \omega _i~
\right]
\int _{-T}^{+T}dt_f~\int _{-T}^{t_f}dt_i~
e^{+i\omega _ft_f}
e^{-i\omega _it_i}\cr
&&\times 
\pmatrix{1\cr 1\cr }^T
\tilde {\bf P}(t_f~\vert ~t_i)
\pmatrix{
{+1\over 2i\tilde \Omega }\cr
{-1\over 2i\tilde \Omega }\cr
}.
\ea
\ba
S_{III\leftarrow II}(\omega _f \vert \omega _i)&=&
{1\over 2\pi }
(-)\left[ {\gamma ^2\over 4}+\tilde \Omega ^2-\omega _i^2+i\gamma \omega _i~ 
\right]
\int _{+T}^{+\infty }dt_f~\int _{-T}^{+T}dt_i~
e^{+i\omega _ft_f}
e^{-i\omega _it_i}\cr
&&\times 
\pmatrix{1\cr 1\cr }^T
\tilde {\bf P}(t_f~\vert +T)~
{\bf J}^{-1}~
{\bf P}(+T ~\vert ~t_i)~
\pmatrix{
{+1\over 2i\tilde \Omega }\cr
{-1\over 2i\tilde \Omega }\cr
}.
\ea
\ba
S_{III\leftarrow III}(\omega _f \vert \omega _i)&=&
{1\over 2\pi }
(-)\left[ {\gamma ^2\over 4}+\Omega ^2-\omega _i^2+i\gamma \omega _i~
\right]
\int _{+T}^{+\infty }dt_f~\int _{+T}^{t_f}dt_i~
e^{+i\omega _ft_f}
e^{-i\omega _it_i}\cr
&&\times 
\pmatrix{1\cr 1\cr }^T
{\bf P}(t_f~\vert ~t_i)
\pmatrix{
{+1\over 2i\Omega }\cr
{-1\over 2i\Omega }\cr
}.
\ea
We next proceed to calculate these six terms more explicitly. 
First, for $S_{I\leftarrow I}$ we may extract and evaluate the integral
\ba
&&
\int _{-\infty }^{-T} dt_f~
\int _{-\infty }^{t_f} dt_i~
e^{+i\omega _ft_f}~
e^{-i\omega _it_i}~
\pmatrix{
e^{(-\gamma/2 +i\Omega )(t_f-t_i)}&0\cr 
0&e^{(-\gamma/2 -i\Omega )(t_f-t_i)}\cr
}
\cr
&=&
\int _{-\infty }^{-T} dt_f~
e^{+i(\omega _f-\omega _i)t_f}
\int _0^{+\infty }d\tau ~e^{+i\omega _i\tau }
\pmatrix{
e^{(-\gamma/2 +i\Omega )\tau }&0\cr
0&e^{(-\gamma/2 -i\Omega )\tau }\cr
}
\cr
&=&
e^{-i(\omega _f-\omega _i)T}~
{(-i)\over (\omega _f-\omega _i)-i\epsilon }
\pmatrix{
{1\over (\gamma/2 -i\Omega -i\omega _i)}&0\cr
0&{1\over (\gamma/2 +i\Omega -i\omega _i)}
}.
\ea
For $S_{II\leftarrow I},$ we may evaluate the two
integrals:
\ba
\int _{-\infty }^{-T}dt_i~e^{-i\omega _it_i}~e^{(-\gamma/2 \pm i\Omega)(-T-t_i)}
=
e^{+i\omega _iT}{1\over \gamma/2 \mp i\Omega -i\omega _i}
\ea
and 
\ba
\int _{-T}^{+T}dt_f~e^{+i\omega _ft_f}~
e^{(-\gamma/2 \pm i\tilde \Omega )(t_f+T)}
=
{e^{-i\omega _fT}-e^{+i\omega _fT}e^{(-\gamma/2 \pm i\tilde \Omega )2T}\over 
\gamma/2 \mp  i\tilde \Omega -i\omega _f}.
\ea
For $S_{III\leftarrow I},$ we have the first integral from
$S_{II\leftarrow I}$ again and also
\ba
\int _{+T}^{+\infty }dt_f~e^{+i\omega _ft_f}~
e^{(-\gamma/2 \pm i\Omega )(t_f-T)}
=
e^{+i\omega _fT}
{1\over \gamma/2 \mp i\Omega -i\omega _f}.
\ea
$S_{II\leftarrow II}$ has the two nested integrals 
\ba
&&
\int _{-T}^{+T}dt_f~
\int _{-T}^{t_f}dt_i~
e^{+i\omega _ft_f}~
e^{-i\omega _it_i}~
e^{(-\gamma/2 \pm i\tilde \Omega )(t_f-t_i)}\cr
&=&
\int _{-T}^{+T}dt_f~
e^{(-\gamma/2 \pm i\tilde \Omega +i\omega _f)t_f}
\int _{-T}^{t_f}dt_i~
e^{(+\gamma/2 \mp i\tilde \Omega -i\omega _i)t_i}
\cr
&=&
\int _{-T}^{+T}dt_f~
e^{(-\gamma/2 \pm i\tilde \Omega +i\omega _f)t_f}~
{
e^{(\gamma/2 \mp i\tilde \Omega -i\omega _i)t_f}
-e^{-(\gamma/2 \mp i\tilde \Omega -i\omega _i)T}
\over \gamma/2 \mp i\tilde \Omega -i\omega _i}
\cr
&=&
{1\over \gamma/2  \mp i\tilde \Omega -i\omega _i}
\left[
{e^{+i(\omega _f-\omega _i)T}
-e^{-i(\omega _f-\omega _i)T} \over i(\omega _f-\omega _i)}
+
{e^{[2(-\gamma/2 \pm i\tilde \Omega )+i(\omega _i+\omega _f)]T}
-e^{+i(\omega _i-\omega _f)T} \over \gamma/2  \mp i\tilde \Omega -i\omega _f}
\right] .
\ea
$S_{III\leftarrow II}$ has the integral
\ba
\int _{-T}^{+T}dt_i~e^{-i\omega _it_i}~
e^{(-\gamma/2 \pm i\tilde \Omega )(-t_i+T)}
=
{e^{-i\omega _iT}-e^{+i\omega _iT}e^{(-\gamma/2 \pm i\tilde \Omega )2T}\over
\gamma/2 \mp  i\tilde \Omega -i\omega _i}.
\ea
$S_{III\leftarrow III}$ has the integral
\ba
&&
\int _{+T}^{+\infty }dt_f~
\int _{+T}^{t_f} dt_i~
e^{+i\omega _ft_f}~ e^{-i\omega _it_i}~
e^{(-\gamma/2 \pm i\Omega )(t_f-t_i)}
\cr
&=&
\int _{+T}^{+\infty }dt_f~
e^{+i(\omega _f-\omega _i)t_f}
\int _0^{t_f-T}d\tau ~e^{+i\omega _i\tau }
e^{(-\gamma/2 \pm i\Omega )\tau }
\cr
&=&
{e^{+i(\omega _f-\omega _i)T}\over \gamma/2 \mp i\Omega-i\omega_i}
 \left[ {i\over (\omega _f-\omega _i)+i\epsilon }
 -{1\over \gamma/2 \mp i\Omega -i\omega _f}\right] .
\ea

We have used the fact that
\ba
\int _0^{+\infty }dt~e^{+i\omega t}={+i\over \omega +i\epsilon },\quad 
\int _{-\infty }^0 dt~e^{+i\omega t}={-i\over \omega -i\epsilon }.
\ea
For further reference, we note that
\ba 
2\pi \delta (\omega )=
(+i)\left({1\over \omega +i\epsilon }-{1\over \omega -i\epsilon }\right),
\qquad P\left( {1\over \omega }\right) ={1\over 2}
\left[ {1\over \omega +i\epsilon }+{1\over \omega -i\epsilon }\right] 
\ea
where $P$ denotes principal part.

The final S matrix takes the form
\ba
S(\omega _f \vert \omega _i)&=&
-2\delta(\omega_f-\omega_i)~ 
{\omega _f^2-\Omega ^2-\gamma ^2/4 -i\gamma \omega _f
\over
\omega _f^2-\Omega ^2-\gamma ^2/4 +i\gamma \omega _f
}\cr
&+&S_{nonsing}(\omega _f \vert \omega _i)
\ea
where the singular 
$\delta $-function term reflects what $S$ would be if
$\Omega =\tilde \Omega $ (i.e., if the string constant
of the oscillator were unchanged) and $S_{nonsing}$
is the smooth part resulting from the Fourier transform
of the wave packet emitted as a result of the change
during the interval $[-T,+T].$ 

The non-singular contribution to the final S matrix is given by
\ba
&-&
\left[
\left( {\omega_i^2 -\tilde \Omega ^2 -\gamma^2/4 -i\gamma\omega_i \over
        \omega_i^2 -\tilde \Omega ^2 -\gamma^2/4 +i\gamma\omega_i}\right)  
-
\left( {\omega_i^2 -\Omega ^2 -\gamma^2/4 -i\gamma\omega_i \over 
        \omega_i^2 -\Omega ^2 -\gamma^2/4 +i\gamma\omega_i}\right)
\right]
{\sin[(\omega_f-\omega_i)T] \over \pi(\omega_f-\omega_i)} \cr
&+&
{1 \over \pi}
{\gamma \omega_i(\Omega ^2-\tilde \Omega ^2)
\over
\Bigl\{ (\omega_i+i\gamma/2)^2-      \Omega^2 \Bigr\}
\Bigl\{ (\omega_i+i\gamma/2)^2-\tilde\Omega^2 \Bigr\}
\Bigl\{ (\omega_f+i\gamma/2)^2-      \Omega^2 \Bigr\}
\Bigl\{ (\omega_f+i\gamma/2)^2-\tilde\Omega^2 \Bigr\} }
\cr
&&\times \Biggl[
\Bigl( \omega_f +\omega_i +i\gamma \Bigr)
  \times \Biggl\{
 -\left[(\omega_f+i{\gamma/2})^2 -\Omega ^2 \right]
      \exp [+i(\omega_f-\omega_i)(-T)] \cr 
&&\quad \quad \quad \quad \quad \quad \quad \quad \quad \quad
  +  \left[(\omega_f+i{\gamma/2})^2 -\tilde\Omega ^2 \right]
      \exp [+i(\omega_f-\omega_i)(+T)]
  \Biggr\} 
\cr 
&&\quad 
+\exp [-\gamma T]~
\exp [+i\omega _f(+T)]~
\exp [-i\omega _i(-T)]~
(\tilde \Omega ^2-\Omega ^2)\cr 
&&\quad
\quad
\times 
\biggl\{ (\omega _f+\omega _i+i\gamma ) \cos [\tilde \Omega (2T)]
-i\Bigl[ (\omega _i+i\gamma /2)(\omega _f+i\gamma /2) + \tilde \Omega ^2\Bigr] 
{\sin [\tilde \Omega (2T)]\over \tilde \Omega }
  \biggr\} 
\Biggr] .
\label{s:matrix}
\ea
The term on the first line arises from the fact that the reflection from the 
oscillator during the interval $[-T,+T]$ occurs with a different phase
from that during its complement when $\vert t\vert >T.$ This term, which approaches a 
delta function as $T\to \infty ,$ assumes that during the interval $[-T,+T]$
the oscillator is described by its asymptotic amplitude and phase 
(as if the change from $\Omega $ to $\tilde \Omega $ had taken place
in the infinite past and there were no transients). This picture, of
course, is an approximation because there were transients, and the remaining
terms give the form of these transients. The two terms on the third 
and fourth lines have the form of scatterers localized in time at 
$t=-T$ and $t=+T,$ respectively. The first is due to the transient
as the oscillator changes its amplitude and phase just after $t=-T$
as a result of the change from $\Omega $ to $\tilde \Omega .$ The 
second term is the same at $t=+T$ as the oscillator frequency changes
back from $\tilde \Omega $ to $\Omega $ under the assumption that
the previous initial transient has completely decayed away. Finally,
the terms on the last two lines result from the fact that the 
first transient from $t=-T$ has not completely decayed away by 
the instant $t=+T.$

The transformations just calculated may be cast into a more 
general framework. What we have just been doing was calculating
the matrix elements of a linear symplectic transformation 
relating the description of the system in terms of ``in''
modes to an equivalent description in terms of ``out''
modes. The two descriptions are related by a linear transformation,
which we may express more abstractly as 
\ba
\pmatrix{
\hat a_{out}(\omega )\cr \hat a_{out}^\dagger (\omega )
}
=
\int _0^\infty  d\omega '
\pmatrix{
S_{++}(\omega \vert \omega ')&
S_{+-}(\omega \vert \omega ')\cr
S_{-+}(\omega \vert \omega ')&
S_{--}(\omega \vert \omega ')\cr
}
\pmatrix{
\hat a_{in}(\omega ')\cr \hat a_{in}^\dagger (\omega ')
}.
\label{sym_trans}
\ea
We know that both the ``in'' and the ``out'' operators
must satisfy the canonical commutation relations. In 
particular, for the ``in'' operators, we have
\ba
[\hat a_{in}(\omega ), \hat a_{in}^\dagger (\omega ')]&=&
\delta (\omega -\omega '),\cr 
[\hat a_{in}(\omega ), \hat a_{in}(\omega ')]&=&0,
\label{c_in}
\ea
where $\omega ,\omega '\ge 0,$ and similarly for
the ``out'' operators 
\ba
[\hat a_{out}(\omega ), \hat a_{out}^\dagger (\omega ')]&=&
\delta (\omega -\omega '),\cr 
[\hat a_{out}(\omega ), \hat a_{out}(\omega ')]&=&0.
\label{c_out}
\ea
However, the linear transformation (\ref{sym_trans}) allows
us to calculate the ``out'' commutation relation using 
eqn.~(\ref{c_in}). The requirement that the
commutation relations thus obtained agree
with those in (\ref{c_out}) leads to a number of 
consistency conditions for the coefficients in (\ref{sym_trans}),
which we give below.

By taking Hermitian conjugates, we obtain that
$S_{\alpha \beta }(\omega \vert \omega ')$ satisfies the conditions:
\ba
S_{++}(\omega \vert \omega ')&=&S_{--}^*(\omega \vert \omega '),\cr 
S_{+-}(\omega \vert \omega ')&=&S_{-+}^*(\omega \vert \omega ').
\ea
Eqn.~(\ref{c_out}) gives 
\ba
\int _0^\infty d\omega ^{\prime \prime }~
\left[
S_{++}(\omega \vert \omega ^{\prime \prime })
S_{++}^*(\omega '\vert \omega ^{\prime \prime })
-
S_{+-}(\omega \vert \omega ^{\prime \prime })
S_{+-}^*(\omega '\vert \omega ^{\prime \prime })
\right]
=\delta (\omega -\omega ')
\ea
and 
\ba
\int _0^\infty d\omega ^{\prime \prime }~
\left[
S_{++}(\omega \vert \omega ^{\prime \prime })
S_{+-}(\omega '\vert \omega ^{\prime \prime })
-
S_{+-}(\omega \vert \omega ^{\prime \prime })
S_{++}(\omega '\vert \omega ^{\prime \prime })
\right]
=0.
\ea
These conditions are necessary and sufficient.

For the ``out" states or quanta, we may
construct a number density operator 
$\hat N_{out}(\omega)=
\hat a_{out}^\dagger(\omega )\hat a_{out}(\omega )$ 
indicating the density of the
number of particles of frequency $\omega$ in the ``out'' space. We find that 
the expectation value of $\hat N_{out}(\omega)$ for the ``in'' vacuum state is
\ba
\langle 0_{in} \vert \hat N_{out}(\omega) \vert 0_{in}\rangle =
\int _0^\infty d\omega '|S_{+-}(\omega \vert \omega ')|^2.
\ea
It follows that the ``in'' and ``out'' vacua will be different as long as
\ba
\langle N_{out}\rangle =\int _0^\infty d\omega~\langle N_{out}(\omega )\rangle 
=\int _0^\infty d\omega~ \int _0^\infty d\omega '
|S_{+-}(\omega \vert \omega ')|^2 >0,
\ea
that is, as long as the expectation value of the operator 
$\hat N_{out}=\int d\omega \hat N_{out}(\omega)$ for the
total number of ``out'' particles created from the ``in'' vacuum state is
different from zero. We will obtain a vanishing expectation
value, hence coinciding vacua, in the case when the string-oscillator
coupling remains constant. 

\subsection{A pair of harmonic oscillators with time dependent parameters}

In this section, generalizing the techniques of the example, we solve the 
equations for a pair of coupled harmonic oscillators with a forcing
term included
\ba
m \left[ {d^2\over dt^2} +
\pmatrix{ 
\omega _1 ^2(t) & \alpha ^2(t) \cr
\alpha ^2(t) & \omega _2 ^2(t) 
}
\right]
\pmatrix{
q _1(t) \cr
q _2(t)
} =
\pmatrix{
F _1(t) \cr
F _2(t)
}.
\ea 
Here $\alpha ^2(t)$ is the coupling between the oscillators, which we
shall make approach to zero sufficiently fast as $|t| \to 0$. We
attach only oscillator 2  to the string so that
\ba
q _1(t)=q_{osc}(t), \quad
q _2(t)=u_{out}(t)+u_{in}(t)
\ea
where the solution on the string is
\ba
u(t,x)=u_{out}(t-x)+u_{in}(t+x)
\ea
for $x \ge 0$. It follows that
\ba
F _1(t) =0, \quad
F _2(t) =\tau \frac{\partial u}{\partial x}\Bigg|_{x=0} 
=\tau \left( 
\frac{\partial u_{in}(t)}{\partial t} 
- \frac{\partial u_{out}(t)}{\partial t}
\right).
\ea
The equation of motion takes the form
\ba
&&
\left[ 
{~d^2\over dt^2}+
\pmatrix{0& 0\cr 0& \gamma \cr }{~d\over dt}
+\pmatrix{\omega _1^2& \alpha ^2(t)\cr 
\alpha ^2(t)& \omega _2^2\cr 
}
\right] 
\pmatrix{q_{osc}(t)\cr u_{out}(t)\cr }\cr
&=&
\pmatrix{D^1(t)\cr D^2(t)\cr}
u_{in}(t)
=\pmatrix{-\alpha ^2(t)\cr 
-{~d^2\over dt^2}+\gamma {~d\over dt}-\omega _2^2\cr }
u_{in}(t),
\ea
where we set
\ba
\alpha ^2(t)=
\left\{
\begin{array}{r@{\quad {\rm if}\quad}l}
\bar \alpha^2 , & \vert t\vert <T,\\
0, & \vert t\vert >T.
\end{array}
\right.
\ea

For $\vert t\vert >T,$ we may write the homogeneous solution in the 
form
\ba
E_{(a)}^i~e^{-\gamma _{(a)} t}
\ea
where
\ba
+i\gamma _{(1)}&=&\omega _{(1)}=+\omega _1,\cr
+i\gamma _{(2)}&=&\omega _{(2)}=-\omega _1,\cr
+i\gamma _{(3)}&=&\omega _{(3)}=+i[\gamma /2+\sqrt{(\gamma /2)^2-\omega _2^2}],
\cr 
+i\gamma _{(4)}&=&\omega _{(4)}=+i[\gamma /2-\sqrt{(\gamma /2)^2-\omega _2^2}]
\ea
and
\ba
E_{(1)}^i=E_{(2)}^i=\pmatrix{1\cr 0\cr },\quad 
E_{(3)}^i=E_{(4)}^i=\pmatrix{0\cr 1\cr }.
\ea
For $\vert t\vert <T,$ the $\tilde \omega _{(\tilde a)}$'s are the
roots $\omega $ of 
\ba
\left| \matrix{
\omega ^2-\omega _1^2 &
-\bar \alpha ^2\cr
-\bar \alpha ^2&
\omega ^2-\omega _2^2+i\gamma \omega \cr 
}
\right|
=0, 
\ea
or equivalently of the quartic polynomial
\ba 
(\omega ^2-\omega _1^2)(\omega ^2-\omega _2^2+i\gamma \omega )
-\bar \alpha ^4=0.
\label{eqn-eleven}
\ea
[Note that for $\gamma \ne ,$ $\bar \alpha \ne 0$ this quartic does not reduce
to a quadratic in $\omega ^2.$] The corresponding eigenvectors are given by
$\tilde E^i_{(\tilde a)},$ so that 
\ba
\tilde E^i_{(\tilde a)}~e^{-\tilde \gamma _{(\tilde a)}t}
\ea
form a complete basis for the homogeneous solutions.
The indices $(\tilde a)$ refer to the basis for $\vert t \vert < T$
whereas the indices $(a)$ refer to the basis for $\vert t \vert > T$.

We define the matching matrix $J_{(\tilde a)(a)}$ so that for all $(a)$ and
$i$ 
\ba
\tilde E^i_{(\tilde a)}&=&\sum _{(\tilde a)}J_{(\tilde a)(a)}~E^i_{(a)},\cr
\tilde \gamma _{(\tilde a)}
\tilde E^i_{(\tilde a)}&=&\sum _{(\tilde a)}J_{(\tilde a)(a)}~
\gamma _{(a)}E^i_{(a)}.
\ea

We define $G^i_{(a)}$ such that for all $i, ~j$ 
\ba
&&\sum _{(a)}G^i_{(a)}~E^j_{(a)}=0,\cr
&&\sum _{(a)}\gamma _{(a)}~G^i_{(a)}~E^j_{(a)}=\delta _{ij},
\ea
and the $\tilde G^i_{(a)}$ are analogously defined.  

Consequently, for $t,$ $t'$ entirely in region I (or entirely in region III)
the Green's function is given by
\ba
G^{ij}(t, t')=E^i_{(a)}~P_{(a)(b)}(t,t')~G^j_{(b)},
\ea
where 
\ba
P_{(a)(b)}(t,t')=\delta _{(a),(b)}~e^{-\gamma _{(a)}(t-t')}=
P_{(a)}(t,t')~\delta _{(a),(b)}
\ea
describes a change of origin, and similarly for points entirely in region II, 
\ba
G^{ij}(t, t')=\tilde E^i_{(\tilde a)}~
\tilde P_{(\tilde a)(\tilde b)}(t,t')~G^j_{(\tilde b)}.
\ea
When points straddle the regions, example for propagation from $I$
into $II$ one obtains 
\ba
G^{ij}(t, t')=\tilde E^i_{(\tilde a)}~\tilde P_{(\tilde a)(\tilde b)}(t,-T)~J_{(\tilde b)(b)}~
P_{(b)(a)}(-T,t')~G^j_{(a)}.
\ea

We now give the expressions for the various Bogoliubov
coefficients.
\ba
S({\rm osc}, +, {\rm out}\leftarrow {\rm osc}, +, {\rm in})=
(J^{-1})_{(a=2)(\tilde a)}~
\tilde P_{(\tilde a)}(+T,-T)~J_{(\tilde a)(b=2)}
\label{osc-osc:1}
\ea
\ba
S({\rm osc}, -, {\rm out}\leftarrow {\rm osc}, +, {\rm in})=
(J^{-1})_{(a=1)(\tilde a)}~
\tilde P_{(\tilde a)}(+T,-T)~J_{(\tilde a)(b=2)}
\label{osc-osc:2}
\ea
\ba
S({\rm osc}, +, {\rm out}\leftarrow {\rm osc}, -, {\rm in})=
(J^{-1})_{(a=2)(\tilde a)}~
\tilde P_{(\tilde a)}(+T,-T)~J_{(\tilde a)(b=1)}
\label{osc-osc:3}
\ea
\ba
S({\rm osc}, -, {\rm out}\leftarrow {\rm osc}, -, {\rm in})=
(J^{-1})_{(a=1)(\tilde a)}~
\tilde P_{(\tilde a)}(+T,-T)~J_{(\tilde a)(b=1)}
\label{osc-osc:4}
\ea
\ba
&&S({\rm osc}, +, {\rm out}\leftarrow \omega _I, {\rm in})=\cr
&&\quad \int _{-\infty }^{-T}dt_I~
(J^{-1})_{(a=2)(\tilde a)}~
 \tilde P_{(\tilde a)}(T,-T)~J_{(\tilde a)(b)}~
 P_{(b)}(-T,t_I)~G^j_{(b)}D^j(t_I)~e^{-i\omega _It_I}\cr
&& \quad +\int _{-T}^{+T}dt_I~
(J^{-1})_{(a=2)(\tilde a)}~
 \tilde P_{(\tilde a)}(+T,t_I)~
 \tilde G^j_{(\tilde a)}D^j(t_I)~e^{-i\omega _It_I}
\label{osc-i:1}
\ea
\ba
&&S({\rm osc}, -, {\rm out}\leftarrow \omega _I, {\rm in})=\cr
&&\quad \int _{-\infty }^{-T}dt_I~
(J^{-1})_{(a=1)(\tilde a)}~
 \tilde P_{(\tilde a)}(+T,-T)~J_{(\tilde a)(b)}~
 P_{(b)}(-T,t_I)~G^j_{(b)}D^j(t_I)~e^{-i\omega _It_I}\cr
&&\quad +\int _{-T}^{+T}dt_I~
(J^{-1})_{(a=1)(\tilde a)}~
 \tilde P_{(\tilde a)}(+T,t_I)~
 \tilde G^j_{(\tilde a)}D^j(t_I)~e^{-i\omega _It_I}
\label{osc-i:2}
\ea
\ba
&&S(\omega _F, {\rm out}\leftarrow \omega _I, {\rm in})=\cr
&&\int _{+T}^{+\infty }dt_F~e^{+i\omega _Ft_F}~
\int _{-\infty }^{-T}dt_I~
 E^{i=2}_{(a)}~P_{(a)}(t_F,+T)~(J^{-1})_{(a)(\tilde a)}~
 \tilde P_{(\tilde a)}(+T,-T)~\cr 
&&\quad \times J_{(\tilde a)(b)}~
 P_{(b)}(-T,t_I)~G^j_{(b)}D^j(t_I)~e^{-i\omega _It_I}\cr
&&+\int _{+T}^{+\infty }dt_F~e^{+i\omega _Ft_F}~
\int _{-T}^{+T}dt_I~
 E^{i=2}_{(a)}~P_{(a)}(t_F,+T)~(J^{-1})_{(a)(\tilde a)}~
 \tilde P_{(\tilde a)}(+T,t_I)~
 \tilde G^j_{(\tilde a)}D^j(t_I)~e^{-i\omega _It_I}\cr
&&+\int _{+T}^{+\infty }dt_F~e^{+i\omega _Ft_F}~
\int _{+T}^{+\infty }dt_I~
 E^{i=2}_{(a)}~P_{(a)}(t_F,+T)~G^j_{(a)}D^j(t_I)~e^{-i\omega _It_I}\cr
&&+\int _{-T}^{+T}dt_F~e^{+i\omega _Ft_F}~
\int _{-\infty }^{-T}dt_I~
 \tilde E^{i=2}_{(\tilde a)}~\tilde P_{(\tilde a)}(t_F,-T)~J_{(\tilde a)(b)}~
 P_{(b)}(-T,t_I)~G^j_{(b)}D^j(t_I)~e^{-i\omega _It_I}\cr
&&+\int _{-T}^{+T}dt_F~e^{+i\omega _Ft_F}~
\int _{-T}^{+T}dt_I~
 \tilde E^{i=2}_{(\tilde a)}~\tilde P_{(\tilde a)}(t_F,t_I)~
 \tilde G^j_{(\tilde a)}D^j(t_I)~e^{-i\omega _It_I}\cr
&&+\int _{-\infty }^{-T}dt_F~e^{+i\omega _Ft_F}~
\int _{-\infty }^{-T}dt_I~
 E^{i=2}_{(a)}~P_{(a)}(t_F,t_I)~G^j_{(a)}D^j(t_I)~e^{-i\omega _It_I} 
\ea
\ba
&&S(\omega _F, {\rm out}\leftarrow {\rm osc}, +, {\rm in})=\cr
&&\quad \int _{+T}^{+\infty }dt_F~e^{+i\omega _Ft_F}~
E^{i=2}_{(a)}~P_{(a)}(t_F,+T)~
 (J^{-1})_{(a)(\tilde a)}~\tilde P_{(\tilde a)}(+T,-T)~
 J_{(\tilde a)(b=2)}\cr
&&\quad +\int _{-T}^{+T}dt_F~e^{+i\omega _Ft_F}~
\tilde E^{i=2}_{(\tilde a)}~
 \tilde P_{(\tilde a)}(t_F,-T)~J_{(\tilde a)(b=2)}
\label{f-osc:1}
\ea
\ba
&&S(\omega _F, {\rm out}\leftarrow {\rm osc}, -, {\rm in})=\cr
&&\quad \int _{+T}^{+\infty }dt_F~e^{+i\omega _Ft_F}~
E^{i=2}_{(a)}~P_{(a)}(t_F,+T)~
 (J^{-1})_{(a)(\tilde a)}~\tilde P_{(\tilde a)}(+T,-T)~
 J_{(\tilde a)(b=1)}\cr
&&\quad +\int _{-T}^{+T}dt_F~e^{+i\omega _Ft_F}~
\tilde E^{i=2}_{(\tilde a)}~
 \tilde P_{(\tilde a)}(t_F,-T)~J_{(\tilde a)(b=1)}.
\label{f-osc:2}
\ea
For the string mode, a positive frequency for $\omega_I, \omega_F$
corresponds to the annihilation operators, a negative frequency
corresponds to creation operators.
In the above formulae, summation over repeated indices is implied, and
in some cases indices are repeated three times because intervening
Kronecker deltas have been suppressed.

The Bogoliubov coefficients connecting the oscillator to itself,
as given by eqns.(\ref{osc-osc:1})--(\ref{osc-osc:4}), can be written as
\ba
S({\rm osc}, S_{out}\leftarrow {\rm osc}, S_{in})&=&
[\delta _{S_{out},+}\delta _{2,(a)}+\delta _{S_{out},-}\delta _{1,(a)}]~
[\delta _{S_{in},+}\delta _{2,(b)}+\delta _{S_{in},-}\delta _{1,(b)}]\cr
&&\quad (J^{-1})_{(a)(\tilde c)}~
e^{-\gamma_{(\tilde c)}(2T)}~ J_{(\tilde c)(b)}
\label{s-o-o}
\ea
where $S_{in}, S_{out}=+,-.$ For large $T,$ the sum over $(\tilde c)$
is dominated by the mode that decays most slowly among the 
eigenvalues of eqn.~(\ref{eqn-eleven}), which are in general non-degenerate.
We shall denote this mode as $(\tilde c_D)$ and its eigenvalue as 
$\Gamma,$ so that in this case eqn.~({\ref{s-o-o}) above 
may be approximated as 
\ba
[\delta _{S_{out},+}\delta _{2,(a)}+\delta _{S_{out},-}\delta _{1,(a)}]~
[\delta _{S_{in},+}\delta _{2,(b)}+\delta _{S_{in},-}\delta _{1,(b)}]~
(J^{-1})_{(a)(\tilde c_D)}~ 
J_{(\tilde c_D)(b)}~e^{-\Gamma (2T)},
\ea
which approaches zero exponentially fast for large $T.$
This attenuation can be interpreted as the decay of the oscillator excitations
into the bulk. When $(\Gamma T)\gg 1,$ any initial quantum
information about the initial state of the oscillator becomes
almost completely dissipated into the bulk. An initial state
expressible as a tensor product of  
oscillator and bulk states becomes highly entangled. 
This means that if today we measure the state of the 
oscillator by recourse to operators acting only on 
the oscillator in terms of the initial state, we are almost
exclusively measuring the correlations of the bulk modes---that
linear combination that subsequently scatters and excites the 
oscillator. If the measurement is instead performed just before
decoupling, the above holds with the straightforward modification
that the second $J$ matrix is suppressed. 

Oscillator excitations caused by incoming
bulk excitations are described by the Bogoliubov coefficients 
in eqns.~(\ref{osc-i:1})--(\ref{osc-i:2}). 
We can rewrite these coefficients as 
\ba
&&S({\rm osc}, S_{out}\leftarrow \omega _I, {\rm in})\cr
&=&
[\delta _{S_{out},+}\delta _{2,(a)}+\delta _{S_{out},-}\delta _{1,(a)}]~
(J^{-1})_{(a)(\tilde c)}~
e^{-\gamma_{(\tilde c)}(2T)}~ J_{(\tilde c)(b)} 
\int _{-\infty }^{-T}dt_I~
e^{-\gamma_{(b)}(-T-t_I)}~
G^{j=2}_{(b)}D^{j=2}(t_I)~ e^{-i\omega _It_I}\cr
&+& 
[\delta _{S_{out},+}\delta _{2,(a)}+\delta _{S_{out},-}\delta _{1,(a)}]~
(J^{-1})_{(a)(\tilde c)} 
\int _{-T}^{+T}dt_I~
e^{-\gamma_{(\tilde c)}(T-t_I)}~
\tilde G^j_{(\tilde c)}D^j(t_I)~ e^{-i\omega _It_I}
\ea
since $D^{j=1}(t)=-\alpha(t)=0$ for $t<-T.$ 
Computing the integrals we obtain
\ba
&&S({\rm osc}, S_{out}\leftarrow \omega _I, {\rm in}) =
[\delta _{S_{out},+}\delta _{2,(a)}+\delta _{S_{out},-}\delta _{1,(a)}]~ 
(J^{-1})_{(a)(\tilde c)}\cr
&&\quad \Bigg(
{e^{-\gamma_{(\tilde c)}(2T)}~
  e^{i\omega _IT} \over {\gamma_{(b)}-i\omega _I} } 
~J_{(\tilde c)(b)}~ G^{j=2}_{(b)}D^{j=2}
+ 
{ {e^{-i\omega _IT}
  -e^{-\gamma_{(\tilde c)}(2T)}e^{i\omega _IT} }
\over {\gamma_{(\tilde c)}-i\omega _I} } 
\tilde G^j_{(\tilde c)}D^j~ 
\Bigg) .
\label{s-o-b}
\ea
For large $T,$ we can suppress the exponentially decaying terms, which
reflect the fact that the oscillator modes have not completely decayed 
into bulk excitations. 
We then find that eqn.~(\ref{s-o-b}) can be approximated for large $T$ as
\ba
[\delta _{S_{out},+}\delta _{2,(a)}+\delta _{S_{out},-}\delta _{1,(a)}]~
(J^{-1})_{(a)(\tilde c)}~
\tilde G^j_{(\tilde c)}D^j~ 
{ e^{-i\omega _IT}
  \over {\gamma_{(\tilde c)}-i\omega _I} }.
\ea

The Bogoliubov coefficients describing production of outgoing
bulk excitations generated by oscillator excitations,
eqns.(\ref{f-osc:1})--(\ref{f-osc:2}),
also consist of two contributions: one from $-T$ to $+T$ during
coupling and another transient after $+T.$
We can rewrite as
\ba
&&S(\omega _F, {\rm out}\leftarrow {\rm osc}, S_{in}) \cr 
&=& \int _{+T}^{+\infty}dt_F~
e^{i\omega _Ft_F}~e^{-\gamma_{(a)}(t_F-T)}
E^{i=2}_{(a)}(J^{-1})_{(a)(\tilde c)}~ 
e^{-\gamma_{(\tilde c)}(2T)}~J_{(\tilde c)(b)}~ 
[\delta _{S_{in},+}\delta _{2,(b)}+\delta _{S_{in},-}\delta _{1,(b)}]
\cr &+& 
\int _{-T}^{+T}dt_F~
e^{i\omega _Ft_F}~
e^{-\gamma_{(\tilde c)}(t_F+T)} 
\tilde E^{i=2}_{(\tilde c)}~ J_{\tilde c,b}~
[\delta _{S_{in},+}\delta _{2,(b)}+\delta _{S_{in},-}\delta _{1,(b)}] \cr
&=&
\Bigg(
E^{i=2}_{(a)}~(J^{-1})_{(a)(\tilde c)}~ 
{ e^{-\gamma_{(\tilde c)}(2T)}
   ~e^{i\omega _FT} \over {\gamma_{(a)}-i\omega_F}}
+\tilde E^{i=2}_{(\tilde c)}~
{ { e^{-i\omega _FT} 
  -e^{-\gamma_{(\tilde c)}(2T)}~e^{i\omega _FT} } 
\over {\gamma_{(\tilde c)}-i\omega_F} }\Bigg) 
\cr &&\quad \quad \quad \times
J_{(\tilde c)(b)}~ 
[\delta _{S_{in},+}\delta _{2,(b)}+\delta _{S_{in},-}\delta _{1,(b)}].
\label{s-b-o}
\ea
For very large $T$ the terms suppressed by an exponential factor
$e^{-\gamma_{(\tilde c)}(2T)}$ reflect a correction due to the fact
that the oscillator mode has not completely been converted into bulk modes.
With this correction suppressed, we find for very large $T$ the 
following approximation for eqn.~(\ref{s-b-o})
\ba 
{ e^{-i\omega _FT} \over {\gamma_{\tilde c}-i\omega_F} }
\tilde E^{i=2}_{(\tilde c)}~J_{(\tilde c)(b)}~
[\delta _{S_{in},+}\delta _{2,(b)}+\delta _{S_{in},-}\delta _{1,(b)}]
\ea
which can be interpreted as a resonance of the oscillator for the bulk
excitation produced.

In the general case (compare with the end of the previous section), 
the linear transformation between the ``in'' and the ``out'' operators 
may be expressed as
\ba
&&\pmatrix{
\hat a_{out}({\rm osc})\cr 
\hat a_{out}^\dagger ({\rm osc} )\cr
\hat a_{out}(\omega )\cr 
\hat a_{out}^\dagger (\omega )\cr
}
=
\pmatrix{
S_{++}({\rm osc}\vert {\rm osc})&
S_{+-}({\rm osc}\vert {\rm osc})&
S_{++}({\rm osc}\vert \omega ')&
S_{+-}({\rm osc}\vert \omega ')\cr 
S_{-+}({\rm osc}\vert {\rm osc})&
S_{--}({\rm osc}\vert {\rm osc})&
S_{-+}({\rm osc}\vert \omega ')&
S_{--}({\rm osc}\vert \omega ')\cr 
S_{++}(\omega \vert {\rm osc})&
S_{+-}(\omega \vert {\rm osc})&
S_{++}(\omega \vert \omega ')&
S_{+-}(\omega \vert \omega ')\cr
S_{-+}(\omega \vert {\rm osc})&
S_{--}(\omega \vert {\rm osc})&
S_{-+}(\omega \vert \omega ')&
S_{--}(\omega \vert \omega ')\cr
}
\pmatrix{
\hat a_{in}({\rm osc})\cr 
\hat a_{in}^\dagger ({\rm osc} )\cr
\hat a_{in}(\omega ')\cr
\hat a_{in}^\dagger (\omega ')\cr
}
\cr
&&
\ea
where summation and integration over repeated indices or variables is implied. 
In the case of several discrete oscillators, $({\rm osc})$ is replaced with
an index taking integer values.

The transformation must preserve the commutation relations
\ba
[\hat a_{in}({\rm osc}), \hat a_{in}^\dagger ({\rm osc}) ]&=&1,\cr
[\hat a_{in}(\omega ), \hat a_{in}^\dagger (\omega )]&=&
\delta (\omega -\omega '),
\ea
and 
\ba
[\hat a_{out}({\rm osc}), \hat a_{out}^\dagger ({\rm osc}) ]&=&1,\cr
[\hat a_{out}(\omega ), \hat a_{out}^\dagger (\omega )]&=& \delta (\omega -\omega '),
\ea
which imply that
\ba
S_{++}({\rm osc}\vert {\rm osc})&=&S_{--}^*({\rm osc}\vert {\rm osc})\cr
S_{+-}({\rm osc}\vert {\rm osc})&=&S_{-+}^*({\rm osc}\vert {\rm osc})\cr
S_{++}({\rm osc}\vert \omega ' )&=&S_{--}^*({\rm osc}\vert \omega ')\cr
S_{+-}({\rm osc}\vert \omega ')&=&S_{-+}^*({\rm osc}\vert \omega ')\cr
S_{++}(\omega \vert {\rm osc})&=&S_{--}^*(\omega \vert {\rm osc})\cr
S_{+-}(\omega \vert {\rm osc})&=&S_{-+}^*(\omega \vert {\rm osc})\cr
S_{++}(\omega \vert \omega ')&=&S_{--}^*(\omega \vert \omega ')\cr
S_{+-}(\omega \vert \omega ')&=&S_{-+}^*(\omega \vert \omega ')
\ea
and
\ba
&& S_{++}({\rm osc}\vert {\rm osc})S_{++}^*({\rm osc}\vert {\rm osc})
-   S_{+-}({\rm osc}\vert {\rm osc})S_{+-}^*({\rm osc}\vert {\rm osc})\cr
&&\quad +\int _0^\infty d\omega ^{\prime \prime }~
\left[
S_{++}({\rm osc}\vert \omega ^{\prime \prime } )
S_{++}^*({\rm osc}\vert \omega ^{\prime \prime } )
-
S_{++}({\rm osc}\vert \omega ^{\prime \prime } )
S_{++}^*({\rm osc}\vert \omega ^{\prime \prime } )
\right] =1,
\ea
\ba
&&
S_{++}(\omega \vert {\rm osc})S_{++}^*(\omega '\vert {\rm osc})
-S_{+-}(\omega \vert {\rm osc})S_{+-}^*(\omega '\vert {\rm osc})\cr
&&\quad +
\int _0^\infty d\omega ^{\prime \prime }~
\left[
S_{++}(\omega \vert \omega ^{\prime \prime })
S_{++}^*(\omega '\vert \omega ^{\prime \prime })
-
S_{+-}(\omega \vert \omega ^{\prime \prime })
S_{+-}^*(\omega '\vert \omega ^{\prime \prime })
\right]
=\delta (\omega -\omega ').
\ea

The remaining commutation relations
\ba
[\hat a_{in}({\rm osc}), \hat a_{in}({\rm osc}) ]=0, \quad 
[\hat a_{in}^\dagger({\rm osc}), \hat a_{in}^\dagger({\rm osc}) ]=0,
\ea
\ba
[\hat a_{in}(\omega), \hat a_{in}(\omega) ]=0, \quad
[\hat a_{in}^\dagger(\omega), \hat a_{in}^\dagger(\omega) ]=0, 
\ea
imply that
\ba
[\hat a_{out}({\rm osc}), \hat a_{out} ({\rm osc}) ]=0, \quad 
[\hat a_{out}^\dagger({\rm osc}), \hat a_{out}^\dagger ({\rm osc}) ]=0,
\ea
\ba
[\hat a_{out}(\omega), \hat a_{out}(\omega) ]=0, \quad 
[\hat a_{out}^\dagger(\omega), \hat a_{out}^\dagger(\omega) ]=0, 
\ea
which entail the additional conditions
\ba
&& S_{++}({\rm osc}\vert {\rm osc})S_{+-}({\rm osc}\vert {\rm osc})
-   S_{+-}({\rm osc}\vert {\rm osc})S_{++}({\rm osc}\vert {\rm osc})\cr
&&\quad +\int _0^\infty d\omega ^{\prime \prime }~
\left[
S_{++}({\rm osc}\vert \omega ^{\prime \prime } )
S_{+-}({\rm osc}\vert \omega ^{\prime \prime } )
-
S_{+-}({\rm osc}\vert \omega ^{\prime \prime } )
S_{++}({\rm osc}\vert \omega ^{\prime \prime } )
\right] =0,
\ea
\ba
&&
S_{++}(\omega \vert {\rm osc})S_{+-}(\omega '\vert {\rm osc})
-S_{+-}(\omega \vert {\rm osc})S_{++}(\omega '\vert {\rm osc})\cr
&&\quad +
\int _0^\infty d\omega ^{\prime \prime }~
\left[
S_{++}(\omega \vert \omega ^{\prime \prime })
S_{+-}(\omega '\vert \omega ^{\prime \prime })
-
S_{+-}(\omega \vert \omega ^{\prime \prime })
S_{++}(\omega '\vert \omega ^{\prime \prime })
\right]
=\delta (\omega -\omega ').
\ea

Similarly, we can construct the number operators for quanta in the
``out'' space, namely 
$N_{out}({\rm osc})=\hat a_{out}^\dagger({\rm osc}) \hat a_{out}({\rm osc})$ 
for the number of oscillator quanta and 
$N_{out}(\omega)=\hat a_{out}^\dagger(\omega) \hat a_{out}(\omega)$ 
for the number of bulk quanta of frequency $\omega$. 
We find the corresponding expectation values
\ba
\langle 0_{in} \vert \hat N_{out}({\rm osc}) \vert 0_{in}\rangle &=&
|S_{+-}({\rm osc} \vert {\rm osc})|^2+
\int _0^\infty d\omega '|S_{+-}({\rm osc} \vert \omega ')|^2 \\
\langle 0_{in} \vert \hat N_{out}(\omega) \vert 0_{in}\rangle &=&
|S_{+-}(\omega \vert {\rm osc})|^2+
\int _0^\infty d\omega '|S_{+-}(\omega \vert \omega ')|^2,
\ea
which will vanish if the ``in'' and ``out'' vacua coincide.

In the application to braneworld cosmology, when we observe
the cosmological perturbations today, we will (using the assumption
of Gaussianity) be measuring, in effect, expectation values
of observables quadratic in the creation and annihilation 
operators localized on the brane today, in other words, the ``out"
oscillator operators. Ignoring the additional discrete indices
(which we suppress),
we offer a useful parameterization of the relevant part
of the S matrix. Let $a_{brane,out}$ and
$a_{brane,out}^\dagger $ be the operators of which we
want to take the quadratic expectation value. The S-matrix
expresses these as a linear combination of 
$a_{brane,out}$ and $a_{brane,out}^\dagger ,$ on the one hand,
and of $a_{bulk,in}(\omega )$ and $a_{bulk,in}^\dagger (\omega ),$ 
on the other. The following offers a useful parameterization of
this transformation. Require that 
$A_{brane,in}$ and $A_{bulk,in},$ normalized
such that $
[A_{brane,in},A_{brane,in}^\dagger ]=
[A_{bulk,in},A_{bulk,in}^\dagger ]=1,$ be entirely on the brane
and in the bulk, respectively. Then $a_{brane,out}$ may be 
expressed in terms of these according to one of the three
following possibilities: either
\ba
a_{brane,out}=\cos \theta ~ A_{brane,in}
            + \sin \theta ~ A_{bulk,in}
\ea
where
$0\le \theta \le \pi /2;$ or ,
\ba
a_{brane,out}=\cosh \xi ~ A_{brane,in}
            + \sinh \xi ~ A_{bulk,in}^\dagger
\ea
where $0\le \xi \le +\infty ;$ or \ba
a_{brane,out}=\sinh \xi ~ A_{brane,in}^\dagger
            + \cosh \xi ~ A_{bulk,in}
\ea
where $0\le \xi \le +\infty .$ We observe that the bulk initial
state may have a very important, even dominant, role in
determining what we see on the brane today. $A_{brane,in},$
and its conjugate, may in turn be related to $a_{brane,in}$
by a Bogoliubov transformation far from the identity, and
the same applies to $A_{bulk,in}.$ 

\section{Generalization to Nontrivial Dynamics on the String:
Reflection, Diffraction, and Dispersion}

In the previous section, because the undulations on the string
were described by a massless wave equation, it was possible
to separate $u(t,x)$ on the string into a sum of the form
$u_{in}(t+x)+u_{out}(t-x).$ The left-moving component
$u_{in}(t)$ acted as a source
for the oscillator, which in turn radiated exclusively
into the right-moving $u_{out}$ channel. However, for a massive 
wave equation of the form 
\ba
\left[ 
 {~\partial ^2\over \partial t^2}
-{~\partial ^2\over \partial x^2}
+m^2(x)
\right] u(t,x)=0,
\label{rf:ccc}
\ea
it is no longer possible to separate into
left-moving and right-moving components. If we consider the
$m^2(x)u(t,x)$ term as a perturbation, we see that
initially right-moving outgoing waves 
are scattered back 
onto the oscillator situated on the boundary,
and incident left-moving waves,
rather than striking the oscillator,
are sometimes rescattered back toward infinity by this term.
In the braneworld problem, the curved geometry, due to the fact that
the bulk spacetime is AdS or AdS-like rather than Minkowski, 
introduces such scattering. 
For the time-independent 
problem, all these effects can be treated straightforwardly by decomposing
into Fourier modes of definite frequency. 
Nothing
more is required than solving eqn.~(\ref{rf:ccc}) by separation
of variables with the appropriate boundary conditions, 
following the formalism developed in section II. 
However, for the time-dependent problem,
when no such decomposition is possible, 
it is necessary to solve for these multiple 
reflections in real space rather than in momentum space. 
One possibility would be to treat the $m^2(x)u(t,x)$
term as a perturbation that generates an infinite series
expansion in the number of reflections. However, for
all but some very special cases, this series 
is either divergent or very slowly convergent. 

A well-behaved series may be obtained by noting that
for all frequencies except those near the resonance
$\omega \approx \omega _0,$ there is very little 
interaction between the string and the oscillator. 
Setting
\ba
u_{out}(\omega )=P(\omega )~u_{in}(\omega ),
\ea
we observe that 
\ba
P(\omega )=(-)\cdot
\frac{\omega ^2-\bar \omega ^2-i\gamma \omega }
     {\omega ^2-\bar \omega ^2+i\gamma \omega }.
\ea
Away from the resonance, the end of the string 
acts almost exactly as a Dirichlet boundary 
condition, for which one would have $P(\omega )=-1$ exactly. 
Consequently, to obtain a rapidly convergent series,
it is advantageous as a first approximation 
to solve for the propagation as if there were Dirichlet 
boundary conditions. Then the violation of the boundary
condition for the oscillator becomes
a source from which a perturbative expansion by successive 
approximations can be generated.

More explicitly, for a given incoming 
wave, with Cauchy initial data specified on past null infinity 
${\cal N}^{(-)},$ we first find a continuation 
$u_0(t,x)$ such that the Dirichlet boundary condition 
\ba
u_0(t,x=0)=0
\ea
is satisfied.
We shall call the $u_0(t,x)$ obtained in this way the ``incident wave.''
We then construct a series expansion
\ba
u(t,x)=\sum _{n=0}^{+\infty }u_n(t,x)
\ea
such that 
\ba
\left[ {~d^2\over dt^2}+\bar \omega ^2(t)\right] u(t,x=0)
=\gamma {\partial u\over \partial x}\Bigg|_{x=0}
\label{eqn:qq}
\ea
is satisfied. Here the string-oscillator 
coupling $\gamma $ (equal to $\tau /m$ where $\tau $ is the string
tension) serves as the expansion 
parameter.
We correct for the violation of eqn.~(\ref{eqn:qq}) 
by $u_0(t,x=0)$
by choosing 
boundary data for $u_1$ such that
\ba
\left[ {~d^2\over dt^2}+\bar \omega ^2(t)\right] u_1(t,x=0)
=\gamma {\partial u_0\over \partial x}\Bigg|_{x=0}
\ea
by means of a retarded Green's function for the oscillator.
Then this Dirichlet boundary data 
for $u_1$ on the boundary is used to extend $u_1$ 
to the full domain $x\ge 0$ using the Dirichlet form of the
retarded Green's function in the bulk, in other words
\ba
u_1(t,x)=\int _{-\infty }^tdt' 
{~\partial \over \partial x'}G_D(t,x;t',x'=0)~u_1(t',x'=0).
\ea
This process repeats itself, so that the relations
\ba
\left[ {~d^2\over dt^2}+\bar \omega ^2(t)\right] u_{n+1}(t,x=0)
=\gamma {\partial u_n\over \partial x}\Bigg|_{x=0}
\ea
and
\ba
u_{n+1}(t,x)=\int _{-\infty }^tdt'
{~\partial \over \partial x'}G_D(t,x;t',x'=0)~u_{n+1}(t',x'=0)
\ea
are satisfied.
At the boundary we use the Green's function for the oscillator
to obtain
\ba
u_{n+1}(t, x=0) = \int_{-\infty}^t dt'~G_{osc}(t,t')~\gamma
 {\partial u_{n}\over \partial x'}(t', x'=0).
\ea
In terms of propagators, we may write 
\ba
u(t,x) &=& u_0(t,x) \cr
&+& \gamma \int_{-\infty}^{t} dt'\int _{-\infty }^{t'} dt^{(2)}~
 D_{x'}G_{bulk}(t,x;t',x'=0)~G_{osc}(t';t^{(2)})~
 D_{x^{(2)}}u_0(t^{(2)},x^{(2)}=0)\cr
&+&\gamma ^2 
 \int _{-\infty }^t dt'
 \int _{-\infty }^{t'} dt^{(2)}~
 \int _{-\infty }^{t^{(2)}} dt^{(3)}~
 \int _{-\infty }^{t^{(3)}} dt^{(4)}\cr 
 &&\times  
 D_{x'}G_{bulk}(t,x;t',x'=0)~G_{osc}(t';t^{(2)})~
 D_{x^{(2)}}D_{x^{(3)}}
  G_{bulk}(t^{(2)}, x^{(2)}=0;
   t^{(3)}, x^{(3)}=0) \cr
 &&\times  
 G_{osc}(t^{(3)};t^{(4)})
  D_{x^{(4)}}
   u_0(t^{(4)},x^{(4)} =0) \cr
&+&\ldots \hfill 
\ea
This expansion contains the infinite sum 
\ba
\bar G_{osc}&=&G_{osc} \cr
&+&\gamma G_{osc}(D_{xI}D_{xII}G_{bulk})G_{osc} \cr
&+&\gamma^2 G_{osc}(D_{xI}D_{xII}G_{bulk})G_{osc}(D_{xI}D_{xII}
G_{bulk})G_{osc}\cr
&+&\gamma^3 G_{osc}(D_{xI}D_{xII}G_{bulk})G_{osc}(D_{xI}D_{xII}G_{bulk})G_{osc}
 (D_{xI}D_{xII}G_{bulk})G_{osc}
+\ldots \hfill
\ea
Here the notation $D_{xI}$ and $D_{xII}$ indicates that the spatial derivative
acts on the spatial component of the first and the 
second argument of $G_{bulk},$ respectively.
We may decompose the internal bulk propagator lines into a singular
and a regular parts
\ba
D_{xI}D_{xII}G_{bulk}=K_{bulk}^{(reg)}+K_{bulk}^{(sing)}.
\ea
Consider for example the massless propagator 
\ba
[\partial _t^2-\partial _x^2]^{-1}(t,x;t',x')=
(1/2)~
\theta (t-t')~
\theta \Bigl( (t-t')^2-(x-x')^2\Bigr) ,
\ea
or the uniform mass propagator 
\ba
&&[\partial _t^2-\partial _x^2+m^2]^{-1}(t,x;t',x')= \cr
&& \quad (1/2)~ \theta (t-t')~ \theta \Bigl(\sqrt{(t-t')^2-(x-x')^2}\Bigr)
J_0\Bigl(m\sqrt{(t-t')^2-(x-x')^2}\Bigr).
\ea
These are the propagators for the entire infinite plane ${\cal R}^2.$ 
We form the retarded Dirichlet propagator by using the method of images so that
\ba
G_D(t,x;t',x')=G_{\infty }(t,x;t',x')-G_{\infty }(t,x;t',-x').
\ea
It turns out that the singular (local) part has the form
\ba
K_{bulk}^{(sing)}(t-t')=-\delta '(t-t'),
\ea
no matter what the propagator is, that is, irrespective of the particular
form of the function $m^2(x)$ in eqn.~(\ref{rf:ccc}). By contrast, 
$K_{bulk}^{(reg)}$ depends sensitively on the particular form of
$m^2(x)$ and encodes all the non-locality arising from propagation from
the brane into the bulk and back again. 

\begin{figure}
\begin{center}
\vskip 3in
\epsfxsize=3in
\epsfysize=6in  
\begin{picture}(300,200) 
\put(80,310){$=$}
\put(150,310){$+$}
\put(80,220){$=$}
\put(150,220){$+$}
\put(210,220){$+$}
\put(270,220){$+~\dots$}
\put(80,125){$=$}
\put(150,125){$+$}
\put(210,125){$+$}
\put(270,125){$+~\dots$}
\put(80,30){$=$}
\put(150,30){$+$}
\put(210,30){$+$}
\put(270,30){$+~\dots$}
\put(40,0){\leavevmode\epsfbox{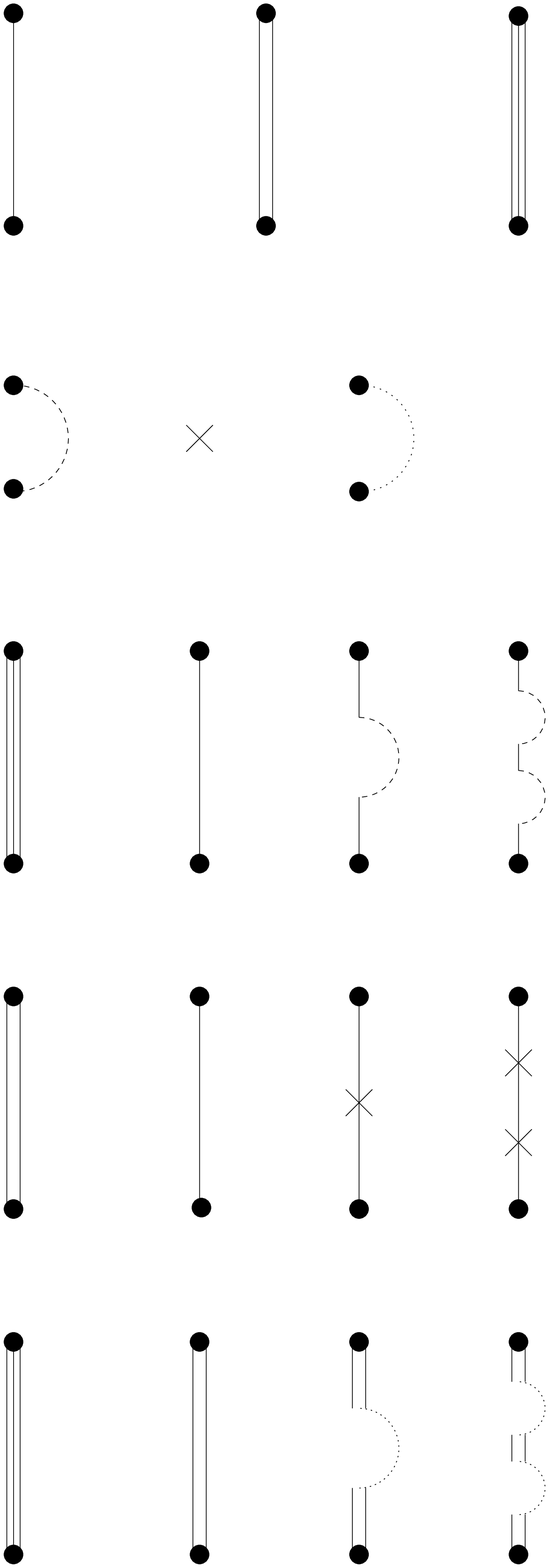}}
\end{picture}
\end{center} 
\vskip 15pt
\caption{\small \baselineskip=4pt {
{\bf Perturbation expansion for the full brane-brane two-point
function as a power series in the brane-bulk coupling $\gamma .$} 
The first line shows three increasingly complete forms
for the brane-brane propagator. The most complete is the 
full propagator, denoted by the triple line, including both
dissipative and nonlocal effects due to the interaction with
the bulk. The double line includes only the dissipative effects
of the bulk interaction, and the single line does not
include any correction at all from the 
brane-bulk interactions. For the simplest case, 
the single line represents  $(D_t^2+\omega _0^2)^{-1}$
and the double line  $(D_t^2+\gamma D_t+\omega _0^2)^{-1}.$
More generally, $\omega _0$ and $\gamma $ may  depend on
time. On the second line we show the diagrammatic decomposition
of the full bulk-bulk propagator $K$ (with both endpoints on the brane
and the brane-bulk coupling factors included), represented
by the dashed curve. This propagator
decomposes into its purely local singular part, 
represented by the cross, and its nonlocal, regular
part, represented by the dotted line. The third line illustrates
the perturbative
expansion for the full brane propagator. On the fourth line
the partial sum over the local insertions is carried out. 
The last line shows the expansion of the full propagator in terms of
the dissipation corrected propagator and the nonlocal corrections.
}}
\label{Fig:feynman}
\end{figure}

\vfill\eject
We use the massless propagator 
\ba
 G_D({\bf \underline{x}}, {\bf \underline{x}}^\prime )
=G_\infty ({\bf \underline{x}}, {\bf \underline{x}}^\prime )
-G_\infty ({\bf \underline{x}}, {\bf \underline{x}}^{\prime R})
\ea 
because the correction due to mass or, more generally, any
potential contributes only to the nonlocal part of $K_{bulk}$
where $t$ and $t'$ do not coincide.

We have 
\ba
K(t,t')&=&\partial _x\partial _{x'} G_D(t,x;t',x')\Bigg\vert _{x=x'=0}\cr
       &=& {1\over 2}
       \lim _{x\to 0+}\left[
       \lim _{x'\to 0+}
       \partial _x\partial _{x'}
       \left(
               \theta (t-t') \cdot
	       \Biggl\{
	          \theta (t-t'+x-x')
	         -\theta (t-t'-x-x')
	       \Biggr\}
       \right)
       \right]
       \cr 
       &=& {1\over 2} \lim _{x\to 0+}\left[
              \lim _{z'\to 0+}
                \Biggl\{
	        -\delta '(t-t'+x-x')
		-\delta '(t-t'-x-x')
	      \Biggr\} 
	      \right] \cr
       &=& -\delta '(t-t').
\ea
The order of the limits is arbitrary; however, by choosing a certain
order as we have done here, we reduce the number of terms to consider.

We now consider the contribution for $t>t'$ where we have used the 
massive propagator
\ba
G_\infty ({\bf \underline{x}}, {\bf \underline{x}}^\prime )=
{1\over 2}\theta (t-t')~\theta (\tau )J_0(m\tau ),
\ea
where
\ba
\tau =\sqrt{(t-t')^2-(z-z')^2}.
\ea
One has 
\ba 
K_{reg}(t-t')&=&{1\over 2}
{~\partial \over \partial x}
{~\partial \over \partial x'}
\Biggl[
 J_0\Bigl( m \sqrt{(t-t')^2-(x-x')^2}\Bigr)
-J_0\Bigl( m \sqrt{(t-t')^2-(x+x')^2}\Bigr)
\Biggr] \cr 
&=& mJ_0^\prime \Bigl( m(t-t')\Bigr) 
{~\partial \over \partial x}
{~\partial \over \partial x'} \sqrt{(t-t')^2-(x-x')^2}\Bigg\vert _{x=x'=0}\cr 
&=& -{mJ_0^\prime \Bigl( m(t-t')\Bigr)\over (t-t')}.
\ea
We observe that the regular part here is in fact regular as $t\to t'+,$
since $J_0'(z)\sim z.$ This is a universal feature.

Since 
\ba 
G_{osc}={1\over D_t^2+\bar \omega ^2(t)},
\label{simp_prop}
\ea
it follows that 
\ba
\hat G_{osc}&=&G_{osc} 
+G_{osc}
\Bigl( \gamma K_{bulk}^{(sing)} \Bigr)
G_{osc} 
+G_{osc}
\Bigl( \gamma K_{bulk}^{(sing)} \Bigr)
G_{osc} 
\Bigl( \gamma K_{bulk}^{(sing)} \Bigr)
G_{osc} 
+\ldots \cr
&=& {1\over {D_t^2+\gamma D_t+\bar \omega ^2(t)}}
\ea
and 
\ba
\bar G_{osc}=
\hat G_{osc}
+\hat G_{osc}
\Bigl( \gamma K_{bulk}^{(reg)}\Bigr)
\hat G_{osc}
+\hat G_{osc}
\Bigl( \gamma K_{bulk}^{(reg)}\Bigr)
\hat G_{osc}
\Bigl( \gamma K_{bulk}^{(reg)}\Bigr)
\hat G_{osc}
+\ldots .
\ea

The above expansions may be interpreted pictorially by means of
Feynman diagrams, as indicated in Fig.~\ref{Fig:feynman}.
The propagator for the oscillator with the coupling to the
bulk ignored given in eqn.~(\ref{simp_prop}) is denoted by a single
vertical line localized on the brane. The dashed curves 
represent propagation of bulk gravitons and their coupling to the brane. 
We decompose the bulk propagator into two parts: a purely local singular part,
denoted by a cross insertion, and a regular, nonlocal
part denoted by a dotted curve, as indicated on the second line 
of the figure. The third line shows how the full 
two-point function on the brane (indicated by the triple line)
may be expressed as an infinite sum. 
The fourth line indicates 
a partial infinite sum  with only the local
insertions included yielding an improved propagator
on the brane (denoted by by the
double vertical line). In terms of the effective equation
on the brane, this infinite summation simply represents the addition
of a local dissipation term to the oscillator equation
of motion. 
Finally, on the last line, an infinite 
sum adding the nonlocality is taken to calculate the 
exact full two-point function on the brane.

\section{Generalization to Moving Boundaries}

For computing braneworld cosmological perturbations, we are 
interested in systems with a moving boundary. In the 
picture where Birkhoff's theorem is exploited so that
the bulk is stationary and the brane moves, except for the 
boundary condition imposed by the moving brane, the propagation
in the bulk is trivial. One simply propagates the bulk gravitons
using  the coordinate system that best exploits the 
symmetries of AdS.

For the scalar wave equations considered here, the 
closest equation to the case of AdS gravitons is
\ba 
\left[ 
{\partial ^2\over \partial t^2}-
{\partial ^2\over \partial x^2}+
k^2+{15/4\over x^2} 
\right] u(t,x)=0,
\ea
where $k$ corresponds to the magnitude of the
wave number in the three 
transverse spatial dimensions. The boundary is
described by a trajectory $x=x_b(t_b),$ which may
be parameterized by proper time $\tau $
where 
\ba
{d\tau \over dt_b}=a(x_b)\cdot \sqrt{1-x_b^{\prime ~2}}
\ea
and $a(x_b)$ is the conformal scalar factor of the metric.
We also define a barred proper time $\bar \tau $ where this
conformal factor has been omitted. 
The oscillator is treated as in the previous section
except that now the proper time $\tau $ takes the place of ordinary 
time. We parameterize $\underline{\bf x}(\tau )=
\Bigl( t_b(\tau ), x_b(\tau )\Bigr) .$
Most everything discussed in the previous section
for the case of a stationary boundary generalizes 
straightforwardly to the case of a moving boundary.
The only slight complication is the computation of the 
Dirichlet Green's functions and the behavior of the $u_{in}$
and $u_{out}$ solutions, which are represented by the external
lines in our diagrammatic expansion. However, this may be 
accomplished by placing a virtual source an infinitesimal
distance just beyond the boundary. This method, which is a sort
of generalization of the method of images, is 
originally due to d'Alembert.\cite{hadamard}

\begin{figure}
\begin{center}
\epsfxsize=3in
\epsfysize=3in
\begin{picture}(300,200)
\put(200,200){$\underline{\bf x}$}
\put(200,15){$\underline{\bf x}'$}
\put(160,100){$\bar \tau$}
\put(80,100){$m\left(\bar \tau,\underline{\bf x}(\bar \tau)\right)$}
\put(80,1){\leavevmode\epsfbox{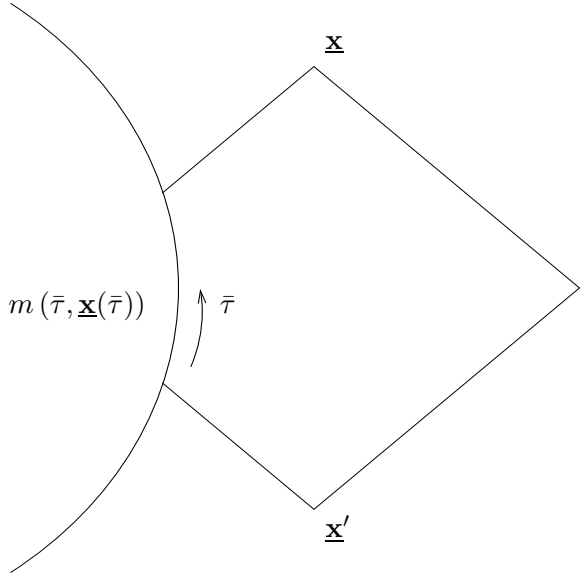}}
\end{picture}
\end{center}
\vskip 15pt
\caption{\small \baselineskip=4pt {
{\bf Method of virtual sources.}
We illustrate how the bulk propagator
from ${\bf \underline{\bf x}'}$ to $\underline{\bf x}$
 may be calculated for a moving boundary
using the method of virtual sources. In general
we do not know the form of the propagator satisfying
the relevant boundary conditions. Therefore, to
compute $G(\underline{\bf x}, {\bf \underline{\bf x}'}),$
we start with an arbitrary propagator, in general violating
the boundary conditions, and then place a virtual source
just infinitesimally beyond the boundary to correct for this
violation. Here a corrective virtual source is required
on that portion of the boundary intersecting with both
the forward lightcone of ${\bf \underline{\bf x}'}$
and the past lightcone of ${\bf \underline{\bf x}},$
which we parameterize by $\bar \tau .$
This virtual source (for the case of Dirichlet
boundary conditions) is a dipole layer, whose
strength as a function of position on the boundary is
determined by solving a Volterra integral equation
of the second kind.
}}
\label{Fig:mov_bd}
\end{figure}

For the external lines in our diagrammatic expansion in the
oscillator/bulk coupling strength $\gamma ,$ we seek two
complete bases of ``in" and ``out" states, characterized by
the asymptotic behavior on past and future null infinity,
respectively. The solutions satisfy 
\ba
{\cal L}_{bulk}u(t,x)=
\left[
{\partial ^2\over \partial t^2}-
{\partial ^2\over \partial x^2}+
k^2 \right]
u(t,x)=0
\ea
where $u$ can be either $u_{in}$ or $u_{out}$ and satisfy
the {\it hard} boundary condition
\ba 
{\cal L}_{B,hard}u=0
\ea
on the boundary $\underline{\bf x}=\underline{\bf x}_b(\bar \tau ).$

By {\it hard} we mean the condition that is local on the boundary
that agrees with the exact boundary condition (which we 
shall call ``soft") in the short-distance or high
frequency limit. For the examples considered in this paper
this is a Dirichlet boundary condition, where
$u=0$ on the boundary, but Neumann and mixed boundary
conditions are allowed, because these do not relate
the behavior at different points on the boundary in a 
nonlocal way. 

For a stationary boundary the asymptotic states
$u_{out}$ and $u_{in}$ may be readily constructed by 
separation of variables.
For a moving boundary accounting for multiple reflections
is less trivial. 

Before calculating the extensions of 
$u_{in}$ and $u_{out}$
based on their behaviors in the neighborhood
and past and future null infinity, respectively,
we first construct the retarded Green's function satisfying
the Dirichlet boundary condition on the null
boundary 
\ba
G_D(\underline{\bf x},\underline{\bf x}')
\ea
starting from the retarded Green's function that
would hold if there were no boundary
\ba
G_\infty (\underline{\bf x},\underline{\bf x}').
\ea

We proceed by postulating a virtual source just
beyond the boundary, propagated according to
$G_\infty ,$ which serves to enforce the boundary
conditions on the moving boundary. For Dirichlet
boundary conditions this is a dipole layer
of strength characterized by the linear
density $m(\bar \tau )$ where the dipole is 
inward and normally directed. The field on the 
boundary of a pointlike dipole of unit strength
at $\tau '$ just beyond the boundary is of the form
\ba 
{1\over 2}\delta (\bar \tau -\bar \tau ')
+\theta (\bar \tau -\bar \tau ')R(\bar \tau ,\bar \tau ')
\ea
where $R(\tau ,\tau ')$ is a smooth, regular function
without any divergence at $\bar \tau =\bar \tau '.$
[For the case of Neumann or mixed boundary conditions, 
we would use a monopole rather than a dipole layer.]

We write as an ansatz 
\ba
G_D(\underline{\bf x},\underline{\bf x}')=G_\infty (\underline{\bf x},
\underline{\bf x}')+
\int _{\bar \tau _a}^{\bar \tau _b}d\bar \tau ~
(\hat n(\bar \tau )\cdot \nabla _2)~
G_\infty \Bigl( \underline{\bf x},\underline{\bf x}(\bar \tau )\Bigr)~
m(\bar \tau ;\underline{\bf x},\underline{\bf x}')
\ea
as indicated in Fig.~\ref{Fig:mov_bd}. 
Here $\hat n(\bar \tau )$ is the inward normal vector on the boundary
and $\nabla _2$ denotes derivative with respect to the second
argument of the Green's function. 
Here the interval $[\tau _a ,\tau _b ],$
which can be empty for certain choices of 
$\underline{\bf x},$ $\underline{\bf x}',$ is the 
intersection of the boundary worldline, the interior of the forward lightcone
of $\underline{\bf x}',$ and the interior of the past lightcone of 
$\underline{\bf x}.$ The virtual
source $m(\bar \tau ; \underline{\bf x},\underline{\bf x}')$ is 
chosen so that the boundary conditions 
are satisfied. This is accomplished by solving the integral equation 
\ba 
{-1\over 2}
m\Bigl( \bar \tau ; 
\underline{\bf x}, \underline{\bf x}'\Bigr)
=G_{\infty }\Bigl( \underline{\bf x}(\bar \tau ),\underline{\bf x}'\Bigr) 
+\int _{\bar \tau _a}^{\bar \tau }d\bar \tau '~
(\hat n(\bar \tau )\cdot \nabla _2)~G_{\infty }\Bigl( \underline{\bf x}(\bar \tau ),
\underline{\bf x}(\bar \tau ')\Bigr) ~
m(\bar \tau ';\underline{\bf x},\underline{\bf x}').
\ea
It is understood that the singular part, which occurs on the left-hand
side of the equation, has been removed from the integral kernel. 
This is a Volterra integral equation of the second kind, which
is always guaranteed to have a unique solution \cite{courant_and_hilbert}.
The ``in" and ``out" functions may be constructed analogously 
using an integral equation.

\section{Discussion}

One of the most interesting questions of braneworld
cosmology is the nature of the cosmological
perturbations predicted and how these might differ
from those possible in models not having a large
extra dimension. In order to carry out the required
calculation, the brane-bulk interaction must be taken
into account, preferably without resorting to 
approximations based on lower-dimensional effective 
descriptions. The techniques developed in this paper,
illustrated using scalar toy models in which a 
(1+1)-dimensional scalar field theory is coupled
to a finite number of degrees of freedom on its
boundary, constitute a step toward achieving
such a complete calculation and demonstrate 
some of the qualitative effects that one might expect.

In section II we studied some simple time-independent
systems. It was seen how coupling the boundary (brane)
degrees of freedom to the continuum (bulk) gave rise
to {\it dissipative} effects from a purely brane point
of view. In addition, lack of spatial homogeneity
in the continuum (as would result from the ``warp
factor" of Randall-Sundrum-like cosmologies) led to {\it nonlocal}
effects, which can be encoded into a spectral density 
$\rho (\omega ).$ In section III we saw
how new effects could arise in time-dependent 
systems (as would arise in an expanding, rather than
static Minkowski, geometry on the brane). The 
discrete degrees of freedom on the brane, rather
than being coupled at the same strength infinitely
far into the past, could initially be uncoupled,
interacting effectively for only a finite amount
of time, so that a sort of $S$ matrix linking 
the ``in" oscillator states to the ``out"
oscillator states for the brane and bulk dynamics
would connect observables today with initial
conditions for the brane and the bulk. When
we observe the braneworld perturbations today,
for the simplest case where these are Gaussian,
it suffices to calculate the expectation values
of operators quadratic in $a_{brane,out}$ and 
$a_{brane,out}^\dagger .$ [Note that our simplified 
notation here assumes 
that the wave number in the three transverse 
spatial dimensions ${\bf k}$ has been fixed
and ignores the fact that these operators also
have an integer index to account for the 
possibility of several modes on the brane.
Similarly, the bulk operators have a discrete
index to account for the several bulk graviton 
polarizations and other possible modes propagating
in the bulk. Since the generalization is straightforward,
we suppress these complication in the notation that 
follows.] In terms
of the ``in" state, $a_{brane,out}$ and
$a_{brane,out}^\dagger $ may be expressed
as a linear combination of ``in" operators,
$A_{brane,in}$ and $A_{bulk,in},$ normalized
so that 
\ba 
[A_{brane,in},A_{brane,in}^\dagger ]=1,
\ea
and 
\ba
[A_{bulk,in},A_{bulk,in}^\dagger ]=1,
\ea
where $A_{brane,in}$ is constructed entirely
as a linear combination of
$a_{brane,in}$ and
$a_{brane,in}^\dagger ,$
and likewise 
$A_{bulk,in}$ is constructed entirely
as a linear combination of
$a_{bulk,in}(\omega )$ and
$a_{bulk,in}^\dagger (\omega ),$
where $\omega \ge 0.$
There are three possibilities: firstly,
\ba
a_{brane,out}=\cos \theta ~ A_{brane,in} 
            + \sin \theta ~ A_{bulk,in}
\ea 
where 
$0\le \theta \le \pi /2;$ secondly,
\ba
a_{brane,out}=\cosh \xi ~ A_{brane,in}
            + \sinh \xi ~ A_{bulk,in}^\dagger 
\ea
where $0\le \xi \le +\infty ;$ and thirdly, 
\ba
a_{brane,out}=\sinh \xi ~ A_{brane,in}^\dagger
            + \cosh \xi ~ A_{bulk,in}
\ea
where $0\le \xi \le +\infty .$ Physically, 
when we measure the perturbations
today, we observe in part the character of
the initial state on the brane and in part
that of the initial state of the bulk. Their
relative importance can be read off from
this S-matrix. 

In section IV we dealt with reflections in the bulk
for the time-dependent situation. Here a wave
initially propagating away from the brane is
scattered, reflected, or diffracted---depending
on which term one prefers---by the bulk so
as to propagate back toward the brane, and vice
versa.  A perturbation expansion was developed 
where ``hard" boundary conditions are imposed
at zeroth order and the ``soft" corrections are
included by successive approximations. A ``hard"
boundary condition is defined as one that is
completely local on the boundary, such as the
Neumann or Dirichlet boundary condition, or a
mixture of the two. These are ``non-dynamical"
because they do not interrelate the boundary behavior
at distinct points on the boundary. By contrast,
``soft" boundary conditions are dynamical and nonlocal,
resulting when the degrees of freedom on the boundary
cede to the incoming waves. Asymptotically, away from
the resonances of the degrees of freedom on the boundary,
the exact ``soft" boundary conditions approach a
corresponding ``hard" boundary condition, which we use
at zeroth order in our perturbation expansion. 

In section V we indicated how to deal with the complication
of a moving boundary. In most respects, the techniques of
the previous sections generalize straightforwardly. The time
for the degrees of freedom on the boundary becomes 
proper time. The principal difficulty in carrying out
the generalization to a moving boundary having an arbitrary 
trajectory (corresponding to an arbitrary expansion history
for braneworld cosmology) is the calculation of the 
bulk Green's functions satisfying the hard boundary 
conditions. This is accomplished by the 
method of virtual sources, where an arbitrary Green's
function (in general not respecting the boundary condition)
is used and then a virtual source placed just infinitesimally 
behind the boundary is postulated in order to emit
into the bulk waves that correct for the violation
of the boundary condition due to using the wrong Green's
function. The source that achieves such a correction
is calculated by solving an integral equation of the 
Volterra form of the second kind.

We are in the process of generalizing this work
to include gravitational gauge fixing so that it can
be applied to calculating braneworld cosmological 
perturbations to linear order. This work will be the 
subject of a future publication.\cite{fut-pub}

{\bf Acknowledgements:} MB thanks Mr Dennis
Avery for supporting this work and the Universit\'e de Paris VII
and the Universit\'e de Paris-Sud for their hospitality during
a visit where part of this work was carried out. 
CC thanks the Programa PRAXIS XXI of the Funda\c c\~ao para a 
Ci\^encia e a Tecnologia and 
the Gulbenkian Foundation for support.
PB thanks the Kavli Institute of Theoretical Physics in 
Santa Barbara for its hospitality. 
We would like to thank Dimitris Angelakis, Cedric Deffayet, Jaume
Garriga, Marc Guilbert, Alan Guth, Valery Rubakov, Dani\`ele Steer, 
Sandip Trivedi, and Neil Turok for useful discussions.

\end{document}